\documentclass{aa}  
\usepackage{hyperref}
\usepackage{natbib}
 \bibpunct{(}{)}{;}{a}{}{,}    
\usepackage{txfonts}
\def\sci{Science}
\def\nata{Nature Astronomy}

\begin{document} 
\title{Lyman-alpha and Lyman continuum emission of extremely metal-poor star-forming galaxies}

\author{Y. I. Izotov\inst{1}
          \and
D. Schaerer\inst{2,3} 
          \and
G. Worseck\inst{4} 
          \and
N. G. Guseva\inst{1}
          \and
J. Chisholm\inst{5,6} 
          \and
T. X. Thuan\inst{7}
          }

\institute{Bogolyubov Institute for Theoretical Physics,
National Academy of Sciences of Ukraine, 14-b Metrolohichna str., Kyiv,
03143, Ukraine\\
\email{yizotov@bitp.kyiv.ua}
\and
Observatoire de Gen\`eve, Universit\'e de Gen\`eve, 
51 Ch. des Maillettes, 1290, Versoix, Switzerland
\and
IRAP/CNRS, 14, Av. E. Belin, 31400 Toulouse, France
\and
VDI/VDE Innovation+Technik GmbH, Steinplatz 1, 10623 Berlin, Germany
\and
Astronomy Department, University of Texas at Austin,
2515 Speedway, Stop C1400 Austin, TX 78712-1205, USA
\and
Cosmic Frontier Center, The University of Texas at Austin, Austin, TX 78712, USA\and
Astronomy Department, University of Virginia, PO Box 400325, Charlottesville,
VA 22904-4325, USA
}

\date{Accepted XXX. Received YYY; in original form ZZZ}

\abstract{
We present observations with the Cosmic Origins Spectrograph onboard the
Hubble Space Telescope of eight compact low-mass star-forming galaxies
selected as the most metal-deficient galaxies from the Sloan Digital Sky Survey
in the redshift range $z$ $\sim$ 0.3 -- 0.4. We aim to extend the
range of metallicities down to 12+log(O/H) $\sim$ 7.5, a range where very few
Lyman Continuum (LyC) emitting galaxies have been studied so far,
and to study the
dependence on metallicity of the Ly$\alpha$ line and Lyman continuum emission.
All selected galaxies are characterised by high production efficiencies of
ionising photons $\xi_{\rm ion}$ = 10$^{25.5}$ -- 10$^{26.0}$ Hz erg$^{-1}$,
high O$_{32}$ ratios of $\sim$5 to $\sim$11 and high equivalent widths
EW(H$\beta$) $\sim$ 130--330\AA,
making them promising candidates to be LyC leaking galaxies.
The Ly$\alpha$ emission line is detected in all eight galaxies with escape fractions
$f_{\rm esc}$(Ly$\alpha$) ranging from $\sim$4\% to $\sim$60\% with the
average value of 27\%. The Lyman continuum in three galaxies is detected at the
level above 3$\sigma$ and in four galaxies at the level below 3$\sigma$.
The LyC in eighth galaxy, J0238+0003, is likely contaminated
by emission from a foreground galaxy. We derive escape fractions
$f_{\rm esc}$(LyC) in the range from $\sim$4\% to high values of up to
$\sim$40\% with the average value of 16\%. 
A strong narrow Ly$\alpha$ emission line with two peaks separated
by $V_{\rm sep}$ $\sim$ 270 -- 404 km s$^{-1}$ was observed in seven galaxies with
detected LyC emission, and very weak Ly$\alpha$ emission with
$V_{\rm sep}$ $\sim$ 712 km s$^{-1}$ is observed in the remaining galaxy with
foreground contamination of Lyman continuum. Addition of new LyC leakers
confirms high scatter of $f_{\rm esc}$(LyC) in previously found
trends on the stellar mass $M_\star$, oxygen
abundance and O$_{32}$, and a tighter correlation of $f_{\rm esc}$(LyC) with $V_{\rm sep}$.
}

\keywords{
(cosmology:) dark ages, reionisation, first stars --- 
galaxies: abundances --- galaxies: dwarf --- galaxies: fundamental parameters 
--- galaxies: ISM --- galaxies: starburst
}

\maketitle
\nolinenumbers

\section{Introduction}\label{intro}

Observations of star-forming galaxies (SFGs) during last decade have revealed
that they can produce copious amount of Lyman continuum (LyC)
emission capable of effectively ionising the intergalactic medium at both
high 
\citep[e.g. ][]{B16,B17,St18,Va18,RT19,W26,Zhu26}
and low redshifts
\citep[e.g. ][]{B14,C17,I16a,I18b,F22a}.
These galaxies are likely analogues of the galaxies at redshifts
6 -- 8, which are considered as the main sources of the reionisation
of the early Universe \citep[e.g. ][]{O09,B15a,Fi19,N20,Je26}.

Direct LyC observations of high-redshift galaxies are difficult because
of their faintness, the increasing of intergalactic medium (IGM) opacity, and
contamination by lower-redshift interlopers
\citep[e.g. ][]{V10,V12,Inoue14,Gr16,Mes25,W26}.
Low-redshift galaxies are brighter, but observations with the
{\sl Hubble Space Telescope} ({\sl HST})/Cosmic Origins Spectrograph
(COS) are needed to detect LyC emission in $z$ $\ga$ 0.3 galaxies.
The Lyman limit in these galaxies is redshifted to wavelengths
$>$ 1120\AA. The {\sl HST} with COS becomes sensitive to emission at these
wavelengths making detection of Lyman continuum possible.
This can be done only for limited samples of low-$z$ galaxies. Due to these
difficulties of direct detection of LyC emission in both the high- and
low-redshift SFGs indirect indicators for the determination of the
LyC escape fraction $f_{\rm esc}$(LyC) can be used.

The shape of the Ly$\alpha$ line can be considered as the prime indicator of the
$f_{\rm esc}$(LyC) value, since it depends on the distribution of the neutral
hydrogen around the galaxy, which also determines the escape of ionising
radiation \citep[e.g. ][]{V15,N22,W26}. In many galaxies with the Ly$\alpha$ emission
line it has a two-peak shape due to scattering in the neutral gas with a
relatively high column density of H~{\sc i}, with a weaker blue peak and a
stronger red peak. The offset of the peaks from the line centre serves
as a measure of the neutral hydrogen optical depth along the line
of sight \citep[e.g. ][]{V15,I18b}.
Other indirect indicators have been considered, 
which use strong emission lines in the rest-frame optical and UV ranges, UV
low-ionisation absorption lines, and far-ultraviolet continuum slope
\citep[e.g. ][]{Ga18,Ga20,Ch18,Ch22,F22a,SL22}.

  \begin{table*}
  \caption{Coordinates, redshifts, distances, equivalent widths EW(H$\beta$),
extinction and some emission-line ratios of selected galaxies
\label{tab1}}
\centering
\begin{tabular}{lccccrc} \hline
  Name&RA(2000.0)&Dec(2000.0)&$z$&$D_L$$^{\rm a}$&$D_A$$^{\rm b}$&EW(H$\beta$)$^{\rm c}$ \\ \hline
J0021$+$3117&00:21:28.99&$+$31:17:26.41&0.3221&1701& 973&181$\pm$18\\ 
J0131$+$2242&01:31:17.80&$+$22:42:09.22&0.3162&1665& 961&135$\pm$8 \\ 
J0238$+$0003&02:38:10.50&$+$00:03:26.68&0.3745&2029&1074&228$\pm$10\\ 
J1247$+$3154&12:47:03.14&$+$31:54:55.09&0.3559&1911&1040&161$\pm$16\\ 
J1340$+$5448&13:40:02.56&$+$54:48:54.81&0.3780&2052&1080&329$\pm$32\\ 
J1607$+$3217&16:07:45.68&$+$32:17:51.96&0.3012&1517& 929&215$\pm$21\\ 
J1638$+$3348&16:38:10.38&$+$33:49:31.09&0.3089&1620& 946&263$\pm$26\\ 
J1658$+$3647&16:58:35.38&$+$36:47:29.36&0.3080&1615& 944&134$\pm$13\\ 
\hline
Name&O$_{32}$$^{\rm d}$&R$_{23}$$^{\rm e}$&log O$_3$$^{\rm f}$&log N$_2$$^{\rm g}$&$A(V)$$^{\rm h}$ \\ \cline{1-6}
J0021$+$3117&10.6$\pm$1.4& 8.5$\pm$0.4&0.769$\pm$0.047&$-$1.755$\pm$0.301&0.21\\
J0131$+$2242& 9.3$\pm$1.0& 7.5$\pm$0.3&0.714$\pm$0.041& ...              &0.17\\
J0238$+$0003& 6.4$\pm$0.5& 8.6$\pm$0.3&0.764$\pm$0.036&$-$1.446$\pm$0.273&0.23\\
J1247$+$3154& 5.5$\pm$0.5& 8.5$\pm$0.3&0.751$\pm$0.045& ...              &0.21\\
J1340$+$5448& 8.0$\pm$1.0& 8.7$\pm$0.4&0.778$\pm$0.055& ...              &0.04\\
J1607$+$3217& 5.5$\pm$0.4& 6.5$\pm$0.2&0.632$\pm$0.027&$-$1.498$\pm$0.270&0.17\\
J1638$+$3348&10.6$\pm$1.6& 7.4$\pm$0.4&0.725$\pm$0.054& ...              &0.11\\
J1658$+$3647& 5.3$\pm$0.6& 8.8$\pm$0.4&0.771$\pm$0.073&$-$1.857$\pm$0.371&0.11\\
\cline{1-6}
\end{tabular}
\tablefoot{$^{\rm a}$Luminosity distance in Mpc \citep[NED, ][]{W06}. $^{\rm b}$Angular size distance in Mpc \citep[NED, ][]{W06}. $^{\rm c}$Equivalent width of the H$\beta$ emission line in \AA. $^{\rm d}$O$_{32}$$\equiv$[O~{\sc iii}]$\lambda$5007/[O~{\sc ii}]$\lambda$3727, where [O~{\sc ii}]$\lambda$3727 denotes the unresolved [O~{\sc ii}]$\lambda$3726+3729 doublet. $^{\rm e}$R$_{23}$$\equiv$([O~{\sc ii}]$\lambda$3727+[O~{\sc iii}]$\lambda$4959+[O~{\sc iii}]$\lambda$5007)/H$\beta$. $^{\rm f}$O$_3$$\equiv$[O~{\sc iii}]$\lambda$5007/H$\beta$. $^{\rm g}$N$_2$$\equiv$[N~{\sc ii}]$\lambda$6583/H$\alpha$. $^{\rm h}$Milky Way extinction in the $V$-band in mags.}

  \end{table*}

\citet{JO13} and \citet{NO14} proposed to use the
O$_{32}$ = [O~{\sc iii}]$\lambda$5007/[O~{\sc ii}]$\lambda$3727 flux ratio.
A high value of this ratio may indicate that the ISM is
predominantly ionised, allowing the escape of Lyman continuum
photons. Indeed, \citet{I16a,I16b,I18a,I18b,I21a} obtained
{\sl HST}/COS observations of compact SFGs at redshifts $z$ $\sim$ 0.3 -- 0.4
with O$_{32}$ = 5 -- 28 and an escape fraction in the range of 2 -- 72
per cent. Although they did find some trend of increasing $f_{\rm esc}$(LyC)
with increasing O$_{32}$, the dependence is weak, with a large scatter.

It has also been suggested that $f_{\rm esc}$(LyC) tends to be higher in
low-mass galaxies \citep{W14}. However, \citet{I18b,I21a}
added lowest-mass LyC leakers and rather found a relatively weak anti-correlation
between $f_{\rm esc}$(LyC) and stellar mass $M_\star$ in a wide range between
10$^7$ -- 10$^{10}$ M$_\odot$. A similar anti-correlation was also found in the
Low-$z$ Lyman Continuum Survey (LzLCS) in \citet{F22b}.

\begin{figure*}
\centering
\includegraphics[angle=0,width=0.90\linewidth]{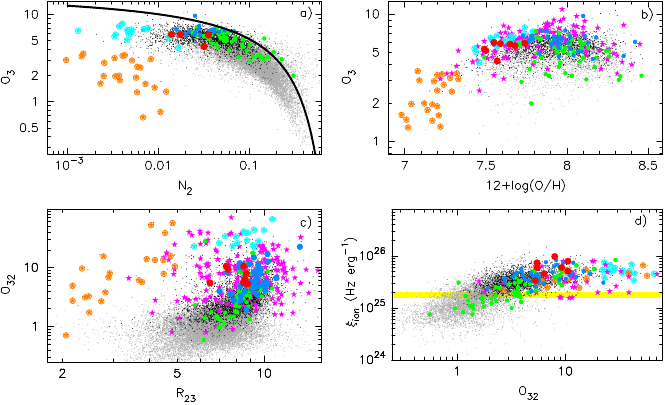}
\caption{a) Diagnostic diagram O$_3$\,=\,[O\,{\sc iii}]5007H$\beta$\,--\,N$_2$\,=\,[N\,{\sc ii}]6584/H$\alpha$. b) Dependence of O$_3$\,=\,[O\,{\sc iii}]5007/H$\beta$ on
oxygen abundance 12\,+\,log(O/H). c)~Diagram
O$_{32}$\,=\,[O\,{\sc iii}]5007/[O\,{\sc ii}]3727\,--\,R$_{23}$\,=\,([O\,{\sc ii}3727\,+\,[O\,{\sc iii}]4959+5007)/H$\beta$, where [O\,{\sc ii}]3727 is
an unresolved doublet [O\,{\sc ii}]3726+3729. d) Dependence of ionising photon
production efficiency $\xi_{\rm ion}$ on O$_{32}$. In all panels, the galaxies
from this paper are shown by red symbols, galaxies from 
\citet{I16a,I16b,I18a,I18b,I21a,I25}, \citet{B14}, \citet{C17}, 
\citet{Xu22} are represented by blue symbols, galaxies from \citet{F22a} are
shown by green symbols, and SDSS galaxies
with EW(H$\beta$) $\geq$ 100\AA\ and EW(H$\beta$) $<$ 100\AA\ are represented by
black and grey dots, respectively. Magenta filled stars in b), c) and d) are
galaxies at $z$ = 6 -- 8 collected by \citet{H26}, \citet{He23}, \citet{Na23},
\citet{Sa24}, \citet{Bo24}, \citet{Ca23}, \citet{RB24}, and \citet{San23}.
Orange encircled stars represent a sample of most metal-deficient galaxies
with 12+log(O/H)$\leq$7.25 at $z$ $<$ 0.1
\citep[][and references therein]{I24b}, whereas cyan encircled stars are
low-redshift galaxies with O$_{32}$ $>$ 20 \citep[][SDSS]{I17,G24}.
Black line in a) by \citet{K03} separates
AGNs and star-forming galaxies. Only SDSS galaxies with [O~{\sc iii}]4363
detected at > 2$\sigma$ are shown in b). Yellow region in d) shows the canonical
value of $\xi_{\rm ion}$ by \citet{R13}. \label{fig1}}
\end{figure*}

  \begin{table*}
  \caption{Integrated characteristics \label{tab2}}
  \begin{tabular}{lcccccccccccc} \hline
Name&12+logO/H$^{\rm a}$&$M_{\rm FUV}$$^{\rm b}$&log$M_\star$$^{\rm c}$ &log$L$(H$\beta$)$^{\rm d}$&$t_b$$^{\rm e}$&SFR$^{\rm f}$&log$\xi_{\rm ion}$$^{\rm g}$&$\alpha$$^{\rm h}$&$r_{50}$$^{\rm i}$&$\Sigma$$_1^{\rm j}$&$\Sigma$$_2^{\rm k}$&$\beta$$^{\rm l}$\\
\hline   
J0021$+$3117&7.59$\pm$0.01&$-$19.39& 7.86    &41.45&3.7& 6.2    &25.90&0.66&0.10&4.5&197.5&$-$2.59\\
J0131$+$2242&7.50$\pm$0.01&$-$19.27& 7.80    &41.98&3.2&21.1\,~~&25.78&0.73&0.11&12.6&555.4&$-$2.54\\
J0238$+$0003&7.65$\pm$0.01&$-$19.38& 8.61    &41.63&3.4& 9.4    &25.66&0.56&0.11&9.5&247.4&$-$2.44\\
J1247$+$3154&7.68$\pm$0.01&$-$19.62& 9.88    &41.73&3.7&11.9\,~~&25.78&0.24&0.23&65.8&71.6&$-$2.51\\
J1340$+$5448&7.55$\pm$0.02&$-$18.87& 8.10    &41.79&0.4& 6.8    &25.99&0.58&0.22&6.4&44.7&$-$2.51\\
J1607$+$3217&7.57$\pm$0.01&$-$19.97&10.23    &41.92&0.6&18.4\,~~&25.87&0.64&0.18&14.3&180.9&$-$2.40\\
J1638$+$3348&7.49$\pm$0.02&$-$18.97& 8.82    &41.30&0.8& 4.4    &25.71& ...$^{\rm m}$& ...$^{\rm m}$& ...$^{\rm m}$&...$^{\rm m}$&$-$2.50\\
J1658$+$3647&7.74$\pm$0.02&$-$19.27& 8.20    &41.33&1.0& 4.7    &25.53&0.52&0.13&5.5&88.6&$-$2.60\\
\hline
  \end{tabular}

\tablefoot{$^{\rm a}$Oxygen abundance derived by the direct $T_{\rm e}$-method. $^{\rm b}$Absolute far-ultraviolet (FUV) magnitude derived from the SED, in mag. $^{\rm c}$$M_\star$ = $M_{\rm y}$ $+$ $M_{\rm o}$ is the total stellar mass in M$_\odot$, where $M_{\rm y}$ and $M_{\rm o}$ are masses of the young and old stellar populations, respectively. $^{\rm d}$$L$(H$\beta$) is the H$\beta$ luminosity corrected for the Milky Way and internal extinction in erg s$^{-1}$. $^{\rm e}$$t_b$ is the starburst age in Myr. $^{\rm f}$Star-formation rate (SFR) corrected for the Milky Way and internal extinction, and escaping LyC radiation in M$_\odot$ yr$^{-1}$. $^{\rm g}$$\xi_{\rm ion}$ is the ionising photon production efficiency corrected for escaping LyC emission, in Hz erg$^{-1}$. $^{\rm h}$Exponential disc scale length in kpc. $^{\rm i}$Galaxy radius, at which near-ultraviolet (NUV) intensity is equal to half of maximal intensity, in kpc. $^{\rm j}$SFR surface density assuming the galaxy radius is equal to $\alpha$ in M$_\odot$ yr$^{-1}$kpc$^{-2}$. $^{\rm k}$SFR surface density assuming the galaxy radius is equal to $r_{50}$ in M$_\odot$ yr$^{-1}$kpc$^{-2}$. $^{\rm l}$$\beta$ is a slope of intrinsic (unattenuated) SED in the rest frame wavelength range of 1500 -- 1700\AA. $^{\rm m}$Acquisition image not obtained.}
  \end{table*}

Finally, the metallicity of the interstellar medium may also be an indicator
of escaping Ly$\alpha$ emission and of escaping
LyC emission because of lower dust content in low-metallicity galaxies. 
Currently, the number of $z$ $\sim$ 0.3 -- 0.4 candidates for LyC leaking
galaxies observed with the {\sl HST}/COS is presently higher than 100
with LyC detections above 3$\sigma$ in more than 30 galaxies
\citep[e.g. ][]{I16a,I16b,I18a,I18b,F22a}. However, the extremely
metal-deficient (XMD) galaxies with conventional 12+log(O/H)~$\leq$~7.65
(corresponding to metallicity Z $\la$ 0.1 Z$_\odot$) \citep{KO00} are very rare.
Even though these are expected metallicities during the epoch of
reionisation \citep{AC22,Cu24,S22,H26}.

The goal of this paper is to determine $f_{\rm esc}$(LyC)
for eight low-mass galaxies with the lowest metallicities available in the
Sloan Digital Sky Survey (SDSS)
at redshift of 0.3 -- 0.4. The oxygen abundance 12+log(O/H) in six galaxies
is lower than 7.65, and in the two remaining galaxies it is slightly above this value.

The properties of the selected SFGs derived from
observations in the optical range are presented in Section~\ref{sec:integr}. The
{\sl HST} observations and data reduction are described in Section~\ref{sec:obs}.
The surface brightness profiles in the UV range are discussed in
Section~\ref{sec:sbp}. In Section \ref{sec:global}, we compare the {\sl HST}/COS spectra with the
extrapolation of the SEDs modelled with the SDSS spectra to the
UV range. Ly$\alpha$ emission and escaping Lyman continuum emission
are discussed in Section~\ref{sec:lya} together with the corresponding escape
fractions. The indirect indicators of escaping LyC emission are
considered in Section~\ref{sec:ind}. We summarise our findings in
Section~\ref{sec:summary}.

  \begin{table}
  \caption{{\sl HST}/COS observations \label{tab3}}
  \begin{tabular}{lcccc} \hline
\multicolumn{1}{c}{}&\multicolumn{1}{c}{}&\multicolumn{3}{c}{Exposure time (s)} \\
\multicolumn{1}{c}{Name}&\multicolumn{1}{c}{Date}&\multicolumn{3}{c}{(Central wavelength (\AA))} \\ 
    &    &ACQ&G140L&G160M \\ \hline
J0021$+$3117&2022-12-29&2$\times$200     &  5218& 3946\\
            &          &         &(800)&(1533)\\
J0131$+$2242&2023-07-17&2$\times$300     &  4962& 3698\\
            &          &         &(800)&(1533)\\
J0238$+$0003&2023-02-17&2$\times$200     &  4936& 3884\\
            &          &         &(800)&(1589)\\
J1247$+$3154&2025-01-04&2$\times$150     &  4913& 3792\\
            &          &         &(800)&(1589)\\
J1340$+$5448&2023-09-23&2$\times$300     &  6791& 4066\\
            &          &         &(800)&(1589)\\
J1607$+$3217&2023-09-07&2$\times$150     &  4981& 3216$^{\rm a}$\\
            &          &         &(800)&(1533)\\ 
J1638$+$3348&2023-09-30&2$\times$0$^{\rm b}$   &  8026& 3745\\
            &          &         &(800)&(1589)\\ 
J1658$+$3647&2023-09-01&2$\times$300     &  5257& 3786\\
            &          &         &(800)&(1533)\\ 
\hline
\end{tabular}

\tablefoot{$^{\rm a}$Partially executed observation. $^{\rm b}$Failed exposure.}

  \end{table}

\section{Selection of the galaxies and characteristics of the sample}\label{sec:integr}

\subsection{Selection criteria}\label{sec:crit}

A sample of eight local compact low-mass SFGs was selected from the
SDSS. The main selection criteria include
\begin{itemize}
\item a sufficiently high redshift $z$ $>$ 0.3 allowing the observation of Lyman
continuum just below the Lyman limit at rest frame $\lambda$ $<$ 912\AA\
with the {\sl HST}/COS and low-resolution grating G140L;
\item the lowest available metallicity of the SDSS galaxies at
$z$\,$>$\,0.3;
\item compact galaxies with angular sizes on the SDSS images much smaller than
the spectroscopic aperture. This allows to detect all the light from the galaxy
and to compare {\sl HST} UV and SDSS optical data without aperture corrections;
\item isolated galaxies on the SDSS images with no other objects
inside the COS aperture with a diameter of 2\farcs5;
\item sufficiently bright in the near-ultraviolet (NUV) range to allow
direct acquisition of the object;
\item high O$_{32}$ and equivalent width EW(H$\beta$) of the H$\beta$
emission line. In this case the galaxies with
young bursts of star formation are selected with hot and massive stars emitting
a large amount of ionising photons, implying a higher probability of
LyC detection;
\item COS safety conditions must be satisfied meaning that no very
bright objects are located in the region with a diameter of 43\arcsec\ around
the selected galaxy;
\item the galaxy was not previously observed or approved for observations
with the COS G140L grating.
\end{itemize}

Using these criteria we selected eight galaxies with the oxygen abundances
12+log(O/H) in the range 7.49 -- 7.74.
We derived these abundances using
SDSS spectra, direct $T_{\rm e}$-method and equations for ionic and total
element abundances from \citet{I06}. We note that in the local Universe
at $z$ $\la$ 0.1 more metal-poor galaxies do exist \citep[e.g. ][]{I18c}, but
at $z$ $\ga$ 0.3 such galaxies are too faint to be selected for spectroscopic
observations in the course of the SDSS because of their low masses and
luminosities. Even the selected galaxies have faint optical and UV magnitudes
of $\sim$ 21 -- 22 mag (Table~\ref{taba1}), which is near the limit allowing a
definite detection of the Lyman continuum in a relatively short exposures of a
few orbits with the {\sl HST}/COS.

Coordinates, redshifts and characteristics of some emission lines are present
in Table~\ref{tab1}. We note that both the
equivalent width of the H$\beta$ emission line EW(H$\beta$) = 134 -- 329\AA\ and
O$_{32}$ ratio of 5.3~--~10.6 are relatively high indicating the presence of young
starbursts with age $t_b$ of a few Myr and high ionising photon production
efficiency $\xi_{\rm ion}$ (Table~\ref{tab2}).

\subsection{Integrated properties of the sample}\label{sec:integrated}

Some integrated characteristics of selected galaxies are represented in
Table~\ref{tab2}. These characteristics are derived from the observed
emission-line intensities (H$\beta$ luminosity $L$(H$\beta$), star-formation
rate SFR), or from spectral energy distribution (SED) fitting (stellar mass
$M_\star$, efficiency
of ionising photon production $\xi_{\rm ion}$, continuum slope in the UV range
$\beta$ and apparent far-ultraviolet (FUV) magnitudes ($M_{\rm FUV}$)).
In all cases the luminosity distance $D_L$ (Table~\ref{tab1}) was
adopted which is derived for a flat cosmological model with the observed
redshift and the parameters $H_0$ = 67.1 km s$^{-1}$ Mpc$^{-1}$,
$\Omega_\Lambda$ = 0.682, $\Omega_m$ = 0.318 \citep{P14}.

We note that all selected galaxies are dwarf objects characterised by
moderately faint absolute FUV magnitudes of $\sim$ --19 -- --20 mag,
moderately low stellar masses of $\sim$ 10$^{8}$ -- 10$^{10}$ M$_\odot$ but
relatively high star-formation rate of $\sim$ 4 -- 21 M$_\odot$ yr$^{-1}$,
a slope $\beta$ = --2.40 -- --2.60 of the intrinsic (unattenuated) SED in the
rest frame wavelength range of 1500 -- 1700\AA, and an ionising photon production
efficiency $\xi_{\rm ion}$ = $N$(LyC)/$L_\nu$(1500\AA)
$\sim$ 10$^{25.5 - 26.0}$~Hz~erg$^{-1}$, where $N$(LyC) is the LyC production rate
and $L_\nu$(1500\AA) is the monochromatic luminosity in the FUV range, again
indicating ongoing star formation.

Internal interstellar extinction $A$($V$)$_{\rm int}$ has been derived from
the observed decrement of hydrogen emission lines in the SDSS
spectra after correction for the Milky Way extinction with $A$($V$)$_{\rm MW}$
from the NED, adopting the \citet{C89} reddening
law and $R$($V$)$_{\rm int}$ = 2.7 and $R$($V$)$_{\rm MW}$ = 3.1. We find that
because of low metallicity the internal interstellar extinction
$A$($V$)$_{\rm int}$ in all galaxies of $\sim$0.1 mag is low.
 
The SEDs in the optical range are derived using SDSS spectra. 
Fitting includes a sum of both the stellar and nebular continua.
A Monte Carlo $\chi$$^2$ minimisation technique with $\sim$ 100000
realisations was used
\begin{itemize}
\item to fit the continuum in the five rest frame intervals of 3050-3100\AA,
3300-3400\AA, 3600-3650\AA, 5100-5200\AA, and 6600-6700\AA, where the SDSS
spectrum is least noisy and free of absorption and emission lines, and
\item to reproduce the observed H$\beta$
and H$\alpha$ emission-line equivalent widths reduced to the rest frame.
\end{itemize}

We use the same model for SED fitting as in previous similar studies by
\citet{I16a,I16b,I18a,I18b,I21a,I22,I25}. It includes the following ingredients:
\begin{itemize}
\item Star-formation history (SFH) and stellar continuum. For SFH we adopt
a model including an instantaneous burst with an age
randomly varying in the interval 0 -- 10 Myr and older stellar component
continuously formed at larger ages with randomly varying time intervals
between 10 Myr and 10 Gyr. Stellar SEDs of single stellar populations
in the range between 0 and 10 Gyr with evolutionary tracks
of non-rotating stars by \citet{G00} and a combination
of stellar atmosphere models \citep{L97,S92} are used to produce the integrated
SED for each galaxy. For each galaxy, we select a grid of stellar
evolutionary tracks, corresponding to its metallicity.
\item The nebular continuum includes free-free and free-bound
hydrogen and helium emission, and hydrogen two-photon emission \citep{A84}.
It is
added to stellar continuum using the observed H$\beta$ equivalent width, the
temperature, and the density, which are derived from the SDSS spectra.
The fraction of nebular continuum near H$\beta$ is determined by the ratio of
the observed H$\beta$ equivalent width EW(H$\beta$)$_{\rm obs}$ 
to the equivalent width EW(H$\beta$)$_{\rm rec}$ for pure nebular emission.
Depending on the electron temperature in the range
$T_{\rm e}$\,=\,10000\,-\,20000\,K the EW(H$\beta$)$_{\rm rec}$
varies from $\sim$\,1100\,\AA\ to $\sim$\,900\,\AA.
\item The \citet{S55} initial mass function (IMF) is adopted, with a
slope of $-$2.35, upper and lower mass limits $M_{\rm up}$ and $M_{\rm low}$ of
100\,M$_\odot$ and 0.1\,M$_\odot$, respectively.
\end{itemize}

Positions on the Baldwin-Phillips-Terlevich (BPT) diagram \citep{BPT81} of four
selected galaxies with detected [N~{\sc ii}] $\lambda$6584 emission line are
shown in Fig.~\ref{fig1}a. All of them are in the upper part of the diagram
for star-forming SDSS galaxies at $z$ $\sim$ 0.3 -- 0.4 characterised by
high [O~{\sc iii}]$\lambda$5007/H$\beta$ $\sim$ 4--6, comparable to that in
high-redshift galaxies during the epoch of reionisation
\citep[e.g. ][]{He23,Na23,H26}. In addition, they are
located on the lowest-metallicity edge for SDSS compact galaxies
(Fig.~\ref{fig1}b). 

These galaxies on the O$_{32}$ -- R$_{23}$ diagram (Fig.~\ref{fig1}c) are at
the upper O$_{32}$ end of the distribution for SDSS galaxies (grey dots),
similar to $z$ = 6 -- 8 galaxies by
\citet{Ca23}, \citet{E23}, \citet{Na23}, \citet{San23}, \citet{Bo24},
\citet{L24}, \citet{RB24}, \citet{Je26}, but they are
somewhat offset to lower R$_{23}$ because of low metallicity. Finally, all
selected galaxies are on the top in the distribution of ionising photon
production efficiency $\xi_{\rm ion}$ (Fig.~\ref{fig1}d) with the values, which
are similar to those in $z$ = 6 -- 8 galaxies in the same range of absolute
FUV magnitudes \citep[e.g. ][]{Sa24,Bo24}.

We note that more extreme lower-mass galaxies
were detected at lower redshifts with $z$ $<$ 0.1 which locate in diagnostic
diagrams away from the location of $z$ $\sim$ 0.3 -- 0.4 LyC galaxies,
considerably extending the range of physical characteristics compared to that
for LyC galaxies.
For comparison, we show in Fig.~\ref{fig1} two samples of lower redshift
galaxies. One sample (orange encircled stars) includes the
most metal-deficient low-redshift SFGs with 12+log(O/H) $<$ 7.25
\citep[][and references therein]{I24b}, which strongly deviate from the
main-seguence of SFGs (grey dots) in all panels of Fig.~\ref{fig1}.
The second sample consists of galaxies with O$_{32}$ ratios above 20 (cyan
encircled stars) \citep[][SDSS]{I17,G24}. Their oxygen abundances and
[O~{\sc iii}] 5007/H$\beta$
ratios are similar to those in LyC candidate galaxies considered in this paper
(red symbols in Fig.~\ref{fig1}c). On the other hand, other characteristics
of galaxies of these two types are considerably different.
Ly$\alpha$ observations are available for many galaxies
shown by cyan and orange symbols in Fig.~\ref{fig1} \citep[e.g. ][]{I20,I24a}.
In general, peak separations of the Ly$\alpha$ profiles in these galaxies are
small indicating considerable leakage of LyC emission \citep{I20,I24a}.
Unfortunately, similar galaxies at $z$ $\geq$ 0.3 can not be observed for the
direct detection of LyC emission because of their intrinsic faintness.

Considered galaxies with extreme properties are rare in the local Universe,
but they likely were common during reionisation. Their integrated
characteristics, such as stellar masses, absolute FUV magninudes, ionising
photon production efficiencies (Table~\ref{tab2}) are remarkably
similar to those of luminous Ly$\alpha$ emitters at the end of reionisation
\citep[e.g. ][]{Ni26}.

\begin{figure*}
\centering
\includegraphics[angle=0,width=0.99\linewidth]{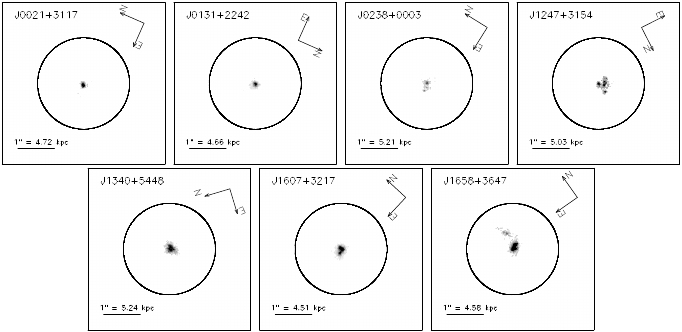}
\caption{The {\sl HST}/COS NUV acquisition images of the candidate LyC leaking
galaxies in a log surface brightness scale. The COS spectroscopic aperture
with a diameter of 2\farcs5 is shown in all panels by a circle.
The linear scale in each panel is derived adopting an angular size distance
$D_A$ (Table~\ref{tab1}).
\label{fig2}}
\end{figure*}

\begin{figure*}
\centering
\includegraphics[angle=0,width=0.99\linewidth]{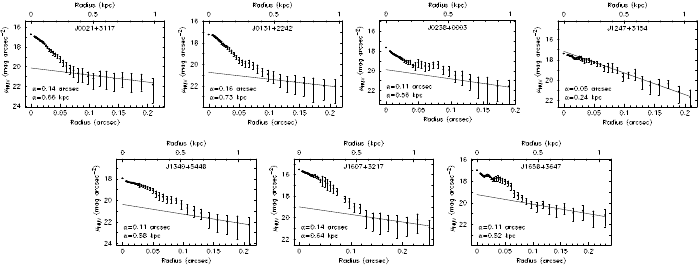}
\caption{NUV surface brightness profiles of galaxies indicated by the dots
  with error bars. Straight lines are linear fits of surface brightness
  $\mu_{\rm NUV}$ in outer galaxy regions, and $\alpha$s are exponential
  scale lengths in arc sec and kpc.
\label{fig3}}
\end{figure*}

\begin{figure*}
\centering
\includegraphics[angle=0,width=0.99\linewidth]{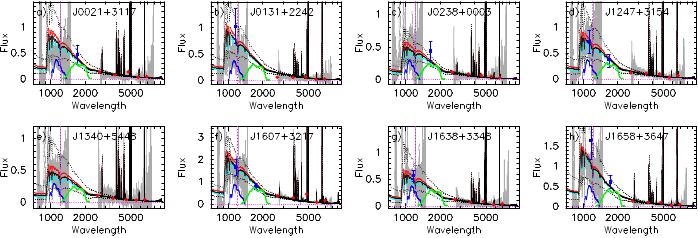}
\caption{A comparison of the rest-frame COS G140L spectra obtained with total
exposures
and SDSS spectra (grey lines), and  photometric data together with the modelled
SEDs of the optical spectra and their extrapolation to the UV range in the
rest frame wavelength scale. {\sl GALEX} FUV and NUV fluxes and SDSS fluxes 
in $u,g,r,i,z$ bands are shown by blue-filled squares with 1$\sigma$
deviations and red-filled 
circles, respectively. Modelled intrinsic SEDs and their extrapolation to
the UV range, which are attenuated by the Milky 
Way extinction with $R(V)_{\rm MW}$ = 3.1 and
internal extinction with $R(V)_{\rm int}$ = 3.1, 2.7 and 2.4, are shown by red,
black and cyan solid lines, respectively. Black dotted lines show the
1$\sigma$ spread of the SED fit reddened with the $C$(H$\beta$)$_{\rm int}$
values and $R(V)_{\rm int}$ = 2.7. Blue and green lines are the
{\sl GALEX} FUV and NUV response curves, respectively. The magenta dotted
vertical and horizontal lines indicate the position of Ly$\alpha$ lines and
zero flux, respectively. Fluxes are in 
10$^{-16}$ erg s$^{-1}$ cm$^{-2}$\AA$^{-1}$, wavelengths are in \AA. \label{fig4}}
\end{figure*}

\section{{\sl HST}/COS observations and data reduction}\label{sec:obs}

{\sl HST}/COS spectroscopy of the eight selected galaxies was obtained
in program GO~17109 (PI: Y.\ I.\ Izotov) during the period December 2022 --
January 2025. The observational details
are presented in Table \ref{tab3}. The galaxies were directly acquired by COS
near ultraviolet (NUV) imaging. All these galaxies are compact on the SDSS
images. The NUV-brightest region of each target was centred in the
2\farcs5 diameter spectroscopic aperture (Fig.~\ref{fig2}).
We note, however, that the acquisition exposure failed for J1638$+$3348
due to guide star acquisition failure. Therefore, the galaxy was blindly
acquired using input coordinates. From the {\sl HST} telemetry, the pointing
offset to the SDSS DR16 position is 0\farcs23, possibly resulting in very
modest wavelength and flux calibration uncertainties for a compact galaxy due
to the offset and partially vignetted COS aperture. 

The spectra were obtained with the 
low-resolution grating G140L with the Lifetime Position LP3 and
medium-resolution grating G160M with the Lifetime Position LP6, applying all 
four focal-plane offset positions.
The 800\,\AA\ setup was used for the G140L grating 
(wavelength sensitivity range of 1100--1950\,\AA, resolving
power $R\simeq 1500$ at 1150\,\AA) to include the redshifted LyC
emission for all targets. We obtained spectra of the galaxies'
Ly$\alpha$ emission lines with the G160M grating ($R$ $\sim$ 15000 at
1600\,\AA), varying the G160M central wavelength with galaxy redshift
to cover the emission line and the nearby continuum on a single detector 
segment. 

The individual exposures were reduced with the {\sc calcos} pipeline v3.6.1 with the
calibration files as of 13 July 2025, followed by accurate (few per cent level)
background subtraction and coaddition with {\sc FaintCOS} v1.2 \citep{Ma21} as
required for our Poisson-limited data. We used the same methods as in
\citet{I18a,I18b,I21a}, but modified the boxcar extraction windows according to
the used COS lifetime position and the grating central wavelength. We used 29
pixel wide G140L boxcar extraction window to account for the wider trace of
the G140L 800\AA\ setup at the longest wavelengths ($\lambda$ $>$ 1700\AA). For
the G160M data a 33 pixel wide boxcar extraction window includes all recorded
Ly$\alpha$ and continuum flux of the targeted galaxies. The G140L (G160M)
spectra were binned by three (four) original pixels to yield a sampling of two
pixels per resolution element. Due to infrequent monitoring of the COS dark
current at the detector voltage used for the G140L observations, we used a
large base set of 139 dark exposures taken during the whole observing period
\citep[see ][for discussion]{I25}. We corrected the G140L data for scattered
geocoronal Ly$\alpha$ emission according to \citet{W16}. We verified its
accuracy by comparing the LyC fluxes obtained in the total exposure and in
orbital night. We find generally good agreement in the measured LyC fluxes,
but due to the often very short observing time in orbital night, we adopt the
total exposures in the following.

\section{Acquisition images and surface brightness profiles in the NUV range}\label{sec:sbp}

The acquisition images of seven galaxies in the NUV range are shown
in Fig.\,\ref{fig2}. All galaxies have angular diameters
considerably smaller than the COS spectroscopic aperture (the circles
in Fig.\,\ref{fig2}) and linear diameters of $\la$ 1 kpc. However, only
two galaxies, J0021+3117 and J0131+2242, have a compact structure with a single
star-forming region. The morphologies of other galaxies are more complex
and diffuse with two or more clumps. However, it is possible that some
clumps are foreground or background objects.
We use these images to derive the surface brightness (SB) profiles
of our galaxies following technique used in previous studies by
\citet{I16b,I18a,I18b,I21a,I22,I25}. 
In accordance with these papers we
found that the outer parts of our galaxies are characterised by a linear
decrease in SB (in mag per square arc sec scale), characteristic of a
disc structure, and by a sharp increase in the central part due to the
bright star-forming region (Fig.\,\ref{fig3}).
The scale lengths $\alpha$ of our galaxies, defined in eq. 1 of
\citet{I16b}, are in the range $\sim$\,0.2\,--\,0.7\,kpc (Fig.\,\ref{fig3}),
somewhat lower
than $\alpha$\,=\,0.6\,--\,1.8\,kpc in other LyC leakers \citep{I16b,I18a,I18b},
but similar to scale lengths of low-mass LyC leakers with
masses $<$\,10$^8$ M$_\odot$ \citep{I21a,I22}. The corresponding surface
densities of star-formation rate in the studied galaxies, $\Sigma$ =
SFR/($\pi\alpha^2$), are similar to those of other LyC leakers. The half-light
radii $r_{50}$ of our galaxies in the NUV are considerably smaller than
$\alpha$ because of the bright compact star-forming regions
in the galaxy centres (see Table\,\ref{tab2}).

\section{Modelled spectral energy distributions in the UV range}\label{sec:global}

To derive the fraction of the escaping ionising radiation we compare
the observed flux in the Lyman continuum and its
intrinsic flux in the same wavelength range \citep[e.g. ][]{I18a}.
The attenuated extrapolations
of SEDs to the UV range along with the observed COS spectra
are shown in Fig.\,\ref{fig4}. For comparison, we also show the
{\sl Galaxy Evolution Explorer} ({\sl GALEX}) FUV
and NUV fluxes with magenta filled squares and the fluxes in the SDSS
$u, g, r, i, z$ filters with blue filled circles. We find that
the spectroscopic and photometric data in the optical range are
consistent, indicating that almost all the emission of our galaxies
is inside the SDSS spectroscopic aperture. Therefore, aperture corrections
are not needed.

The attenuated modelled intrinsic SEDs in the optical range and their
extrapolations to the UV range (Fig.\,\ref{fig4}) are obtained by assuming
that extinctions for stellar and nebular emission are equal and adopting
the extinction coefficients $C$(H$\beta$)$_{\rm MW}$ from the NED and
$C$(H$\beta$)$_{\rm int}$ derived from the hydrogen Balmer decrement
and the reddening law by \citet{C89} at $\lambda$ $\geq$ 1250\AA\ and its
extension to shorter wavelengths by \citet{Ma90} with $R(V)_{\rm int}$ = 3.1
(red solid lines), $R(V)_{\rm int}$ = 2.7 (black solid lines) and $R(V)_{\rm int}$
= 2.4 (cyan solid lines). \citet{Ma90} presents the data only for
$R(V)$ = 3.1. For practical use, we fit them with polynomials and normalised
them to have the same values at $\lambda$=1250\AA\ and for a variety of
$R(V)$s as the values from the \citet{C89} reddening law. The dotted lines
indicate the range of attenuated
SEDs adopting $R(V)_{\rm int}$ = 2.7 and varying $C$(H$\beta$) within 1$\sigma$
errors of its nominal value. The modelled extrapolations of the SEDs to the
UV range and {\sl GALEX} NUV magnitudes are in reasonably good agreement with
the observed COS spectra. On the other hand,
the {\sl GALEX} FUV fluxes in two galaxies, J0131+2242 and J1658+3647, are
systematically above the observed
COS spectra by $\sim$100\% and $\sim$50\%, respectively, in part because
of contribution of strong Ly$\alpha$ emission line
to photometric data. It is seen in Fig.~\ref{fig4} that Ly$\alpha$ in all
galaxies is located near the maximum of the {\sl GALEX} FUV response curve.
Adopting the equivalent widths of Ly$\alpha$ emission line from Table~\ref{tab4}
and {\sl GALEX} FUV band pass equivalent width of 268\AA\ we estimate the
contribution of Ly$\alpha$ of $\sim$45\% in J0131+2242 and $\sim$30\% in
J1658+3647. Similar contribution is expected for other galaxies, whereas no
evident deviation of {\sl GALEX} FUV fluxes from COS fluxes is present in
Fig.~\ref{fig4}. However, we note that uncertainty of {\sl GALEX} FUV photometry
is too high to definitely derive the contribution of Ly$\alpha$ emission to FUV
photometric flux. This problem is not important for SED fitting
because photometric data in both the UV and optical ranges and COS spectra
are not used for that purpose.

  \begin{table*}
  \caption{Parameters for the Ly$\alpha$ emission line \label{tab4}}
\centering
\begin{tabular}{rcrcrrccc} \hline
Name&$A$(Ly$\alpha$)$_{\rm MW}$$^{\rm a}$&\multicolumn{1}{c}{$I$(Ly$\alpha$)$^{\rm b}$}&log $L$(Ly$\alpha$)$^{\rm c}$&\multicolumn{1}{c}{EW(Ly$\alpha$)$^{\rm d}$}&\multicolumn{1}{c}{$V_{\rm sep}$$^{\rm e}$}&\multicolumn{1}{c}{$I$(H$\beta$)$^{\rm f}$}&\multicolumn{1}{c}{$f_{\rm esc}$(Ly$\alpha$)$^{\rm g}$}&\multicolumn{1}{c}{blue/red$^{\rm h}$} \\ 
\hline
J0021$+$3117&0.640& 49.3$\pm$11.9&42.23&168.4$\pm$2.4& 343.8$\pm$40.0& 10.2$\pm$1.2&20.7$\pm$5.4&0.18$\pm$0.04 \\
J0131$+$2242&0.508& 79.4$\pm$8.4&42.42&172.8$\pm$4.9& 270.0$\pm$51.8&  5.5$\pm$0.3 &62.0$\pm$7.2&0.22$\pm$0.03 \\
J0238$+$0003&0.216&  8.2$\pm$3.5&41.61& 21.5$\pm$6.1& 712.5$\pm$41.0&  9.7$\pm$0.5 &3.6$\pm$1.5&0.72$\pm$0.19 \\
J1247$+$3154&0.108& 31.7$\pm$9.9&42.14&127.7$\pm$6.6& 404.0$\pm$61.2& 13.5$\pm$0.8 &10.4$\pm$3.2&0.52$\pm$0.07 \\
J1340$+$5448&0.069& 83.7$\pm$8.8&42.62&197.5$\pm$2.2& 339.3$\pm$58.7&  6.4$\pm$1.4 &56.0$\pm$13.2&0.38$\pm$0.03 \\
J1607$+$3217&0.244&146.6$\pm$10.6&42.61& 87.1$\pm$10.& 377.8$\pm$79.5& 36.4$\pm$1.1 &17.3$\pm$1.3&0.27$\pm$0.02 \\
J1638$+$3348&0.188& 55.3$\pm$7.8&42.24&127.0$\pm$4.5& 319.1$\pm$40.0&  7.1$\pm$0.5 &33.4$\pm$5.1&0.20$\pm$0.03 \\
J1658$+$3647&0.110& 45.7$\pm$5.8&42.15&100.6$\pm$12.& 338.9$\pm$27.6&  7.6$\pm$0.5 &24.6$\pm$3.6&0.16$\pm$0.03 \\ \hline
  \end{tabular}

\tablefoot{$^{\rm a}$Milky Way extinction at the observed wavelength of the Ly$\alpha$ emission line in mags adopting \citet{C89} reddening law with $R(V)$=3.1. $^{\rm b}$Flux in 10$^{-16}$ erg s$^{-1}$ cm$^{-2}$ measured in the COS spectrum and corrected for Milky Way extinction. $^{\rm c}$$L$(Ly$\alpha$) is Ly$\alpha$ luminosity in erg~s$^{-1}$ corrected for Milky Way
extinction. $^{\rm d}$Rest frame equivalent width in \AA. $^{\rm e}$Ly$\alpha$ peak separation in km s$^{-1}$. $^{\rm f}$Flux in 10$^{-16}$ erg s$^{-1}$ cm$^{-2}$ measured in the SDSS spectrum and corrected for extinction derived from the Balmer decrement. $^{\rm g}$Ly$\alpha$ escape fraction
in per cent. $^{\rm h}$Blue-to-red peak flux ratio.}

\end{table*}

\begin{figure*}
\centering
\includegraphics[angle=0,width=0.99\linewidth]{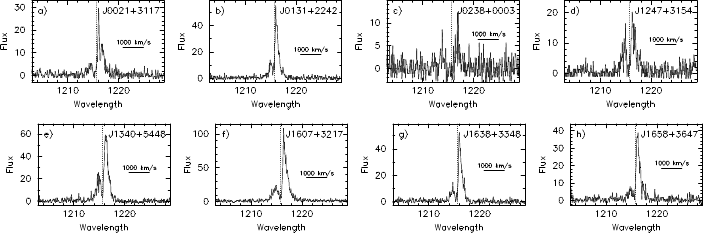}
\caption{Ly$\alpha$ profiles. Vertical dotted lines indicate the rest frame 
wavelength of 1215.67\AA\ for Ly$\alpha$. 
Fluxes are in 10$^{-16}$ erg s$^{-1}$ cm$^{-2}$\AA$^{-1}$ and rest frame 
wavelengths are in \AA. \label{fig5}}
\end{figure*}

\begin{figure*}
\centering
\includegraphics[angle=0,width=0.99\linewidth]{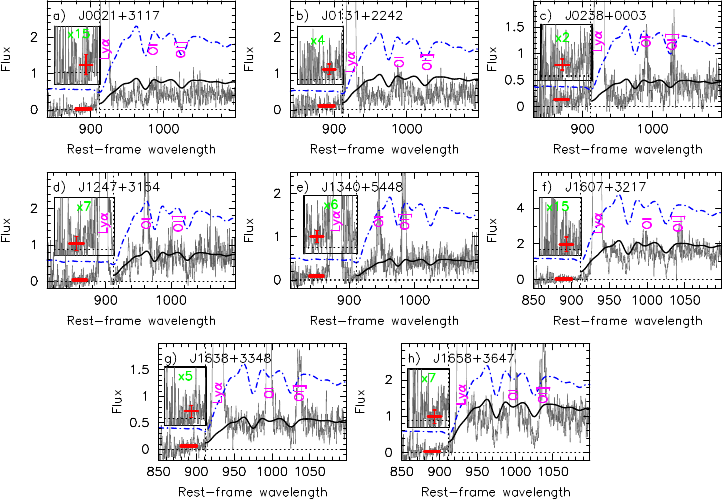}
\caption{Rest frame COS G140L spectra of our sources (grey lines).
The LyC fluxes shown by the solid red horizontal lines are measured in the
wavelength ranges determined by their location.
The extrapolations to the UV
range of intrinsic SEDs and of attenuated SEDs in the optical range
adopting a $R(V)$ = 2.7 are represented by blue dash-dotted lines
and black solid lines, respectively. Expanded versions of the Lyman
continuum with scaling factor along the ordinate (in green) are shown in insets,
  where horizontal red line and error bar are the
average LyC flux in the range determined in Table~\ref{tab5} and its
3$\sigma$, respectively. The Lyman limit at the rest frame
wavelength 912\,\AA\ is indicated by dotted vertical lines in both the
figure and inset and zero flux is represented by dotted horizontal lines.
Positions of airglow emission lines Ly$\alpha$, O{\sc i} $\lambda$1304 and
O~{\sc i}] $\lambda$1365 are labelled in magenta.
Fluxes are in 10$^{-16}$ erg s$^{-1}$ cm$^{-2}$\AA$^{-1}$, wavelengths 
are in \AA. \label{fig6}}
\end{figure*}

  \begin{table*}
  \caption{LyC escape fraction \label{tab5}}
  \centering
\begin{tabular}{lccrrrr} \hline
Name&$\lambda_0$$^{\rm a}$&$A$(LyC)$_{\rm MW}$$^{\rm b}$&$I_{\rm mod}$$^{\rm c,d}$&$I_{\rm obs}$$^{\rm c,e}$&$I_{\rm esc}$$^{\rm c,f}$&\multicolumn{1}{c}{$f_{\rm esc}$$^{\rm g}$} \\
    &(\AA)&(mag)&&&&\multicolumn{1}{c}{(per cent)} \\ 
\hline
\vspace{0.1cm}
J0021$+$3117&880-900&0.838& 54.92$\pm$1.92& 3.21$^{+1.51}_{-1.37}$$^{+0.00}_{-0.13}$& 6.95$^{+3.27}_{-2.96}$$^{+0.00}_{-0.28}$&12.7$^{+5.9}_{-5.4}$$^{+0.0}_{-0.5}$ \\
\vspace{0.1cm}
J0131$+$2242&880-900&0.669& 50.96$\pm$3.70&11.49$^{+2.09}_{-1.84}$$^{+0.00}_{-0.21}$&21.28$^{+3.87}_{-3.41}$$^{+0.00}_{-0.39}$&41.7$^{+5.9}_{-5.3}$$^{+0.0}_{-1.0}$ \\
\vspace{0.1cm}
J0238$+$0003$^{\rm h}$&840-860&0.291& 38.29$\pm$4.87&13.12$^{+2.25}_{-1.98}$$^{+0.00}_{-0.22}$&17.15$^{+2.94}_{-2.59}$$^{+0.00}_{-0.29}$&\multicolumn{1}{c}{...} \\
\vspace{0.1cm}
J1247$+$3154&850-870&0.149& 54.29$\pm$4.87& 4.35$^{+1.89}_{-1.70}$$^{+0.00}_{-0.17}$& 4.99$^{+2.17}_{-1.95}$$^{+0.00}_{-0.20}$& 9.2$^{+4.0}_{-3.6}$$^{+0.0}_{-0.4}$ \\
\vspace{0.1cm}
J1340$+$5448&840-860&0.093& 48.10$\pm$5.30& 8.26$^{+1.54}_{-1.51}$$^{+0.13}_{-0.15}$&9.00$^{+1.68}_{-1.65}$$^{+0.14}_{-0.16}$&18.7$^{+3.9}_{-3.3}$$^{+0.3}_{-0.3}$ \\
\vspace{0.1cm}
J1607$+$3217&880-900&0.324&104.12$\pm$5.85& 3.57$^{+2.00}_{-1.82}$$^{+0.00}_{-0.18}$& 4.81$^{+2.70}_{-2.45}$$^{+0.00}_{-0.24}$& 4.6$^{+2.6}_{-2.3}$$^{+0.0}_{-0.2}$ \\ 
\vspace{0.1cm}
J1638$+$3348&880-900&0.249& 38.07$\pm$6.51& 6.02$^{+1.50}_{-1.67}$$^{+0.19}_{-0.01}$& 7.57$^{+1.88}_{-2.10}$$^{+0.24}_{-0.01}$&19.9$^{+6.5}_{-7.2}$$^{+0.8}_{-0.1}$ \\ 
\vspace{0.1cm}
J1658$+$3647&880-900&0.145& 53.29$\pm$1.87& 3.41$^{+1.59}_{-1.77}$$^{+0.14}_{-0.00}$& 3.90$^{+1.82}_{-2.02}$$^{+0.16}_{-0.00}$& 6.8$^{+3.2}_{-3.5}$$^{+0.3}_{-0.0}$ \\
\hline
\end{tabular}
\tablefoot{$^{\rm a}$Rest frame wavelength range in \AA\ used to determine the LyC flux. $^{\rm b}$Milky Way extinction at the mean observed wavelengths of the 
  range used to determine the LyC flux. The \citet{C89} reddening law with $R(V)$ = 3.1 is adopted. $^{\rm c}$in units of 10$^{-18}$ erg s$^{-1}$cm$^{-2}$\AA$^{-1}$. $^{\rm d}$LyC flux derived from extrapolation of the modelled SED in the optical range to the UV range. $^{\rm e}$Observed LyC flux. $^{\rm f}$LyC flux which is corrected for the Milky Way extinction. $^{\rm g}$$f_{\rm esc}$(LyC) = $I_{\rm esc}$/$I_{\rm mod}$, where $I_{\rm mod}$ is derived from SED. $^{\rm h}$Excluded from LyC determination because of possible contamination.
}

\end{table*}

It is seen in Fig.\,\ref{fig4} that the SDSS spectra are reproduced by the
models quite well. On average, extrapolations of the attenuated SEDs
to the UV range with $R(V)_{\rm int}$\,=\,2.7 and 2.4.
reproduce the observed COS spectra somewhat better with flux deviations not
exceeding $\sim$ 10 per cent for all galaxies, including J1638+3348, the object
with failed acquisition exposure. 

We note that
Fig.\,\ref{fig4} is used only for the sake of illustration to check whether
extrapolation of the SED in the optical range reproduces the observed COS
spectrum. But it is not used for the determination of the escaping LyC fraction.
Instead, the observed LyC flux is measured in COS spectra and the intrinsic LyC
flux is determined from simultaneous fitting of the SED in the
optical range (Fig.\,\ref{fig4}) and of observed equivalent
widths of the H$\beta$ and H$\alpha$ emission lines (Table\,\ref{taba2}).
The fluxes of latter lines are also iteratively corrected for
the escaping ionising radiation \citep[e.g. ][]{I18b} and they determine the
level of the intrinsic LyC emission.
It is seen in Fig.\,\ref{fig4} that the SED in the optical range is
almost independent on $R(V)_{\rm int}$. Consequently, the LyC escape fraction
$f_{\rm esc}$(LyC) is also almost independent of
$R(V)_{\rm int}$. This is because $f_{\rm esc}$(LyC) is derived
from the ratio of the observed to modelled intrinsic LyC fluxes with the latter
fluxes being derived from data in the optical range.

\begin{figure*}
\centering
\includegraphics[angle=0,width=0.99\linewidth]{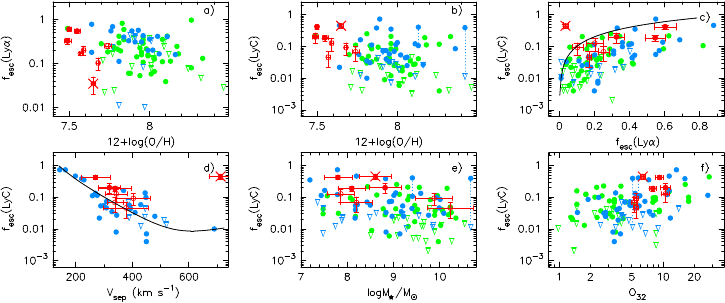}
\caption{a) and b) Relations between the Ly$\alpha$ and Lyman
continuum escape fractions
$f_{\rm esc}$(Ly$\alpha$) and $f_{\rm esc}$(LyC) in low-redshift LyC leaking
galaxies and oxygen abundance 12+log(O/H). c) Relation
between $f_{\rm esc}$(LyC) and $f_{\rm esc}$(Ly$\alpha$). d) Relation
between $f_{\rm esc}$(LyC) and peak separation $V_{\rm sep}$ of the
Ly$\alpha$ emission line. e) Relation between $f_{\rm esc}$(LyC)
and stellar mass $M_\star$. f) Relation between $f_{\rm esc}$(LyC)
and O$_{32}$ ratio. The LyC escape fraction $f_{\rm esc}$(LyC) in all panels
is derived by the
method with the use of SED fits and constraints from the observed H$\beta$ and
H$\alpha$ equivalent widths. Red filled and open circles are galaxies
studied in this paper with LyC detection at $>$3$\sigma$ and $<$3$\sigma$,
respectively. In all panels, galaxies shown in blue and green are the same as
in Fig.~\ref{fig1}, but filled and open symbols indicate galaxies with
definite LyC detection and its upper limits, respectively.
The galaxy J0238+0003 with likely foreground contamination is
crossed in all panels. The solid line in c) is the equality line 
and the solid line in d) represents the relation from \citet{I18b}.
\label{fig7}}
\end{figure*}

\section{Ly$\alpha$ and LyC emission}\label{sec:lya}

A resolved Ly$\alpha$ $\lambda$1216\,\AA\ emission line is detected
in the G160M medium-resolution spectra of all eight galaxies (Fig.\,\ref{fig5}).
Its double-peaked profile is similar to that observed in most
known LyC leakers, in particular, in $z$ $\sim$ 0.3 leakers \citep[e.g. ][]{I16b,I18a,I18b,I21a,I22}, and in some other
galaxies at lower redshift \citep[e.g. ][]{JO14,H15,Y17,I20,I24a}.
Using these
observations we derived some characteristics of the Ly$\alpha$ emission line
in all galaxies, such as the rest-frame Ly$\alpha$ flux and equivalent width,
both measured for the entire line, the peak separation $V_{\rm sep}$ and the flux
ratio of blue-to-red components. We note that uncertainties of these
characteristics are determined by statistical errors. On the other hand, some
other characteristics, not considered here, e.g. Ly$\alpha$ offset from
systemic velocity, are additionally affected by uncertainties of wavelength
calibration. We estimate these uncertainties to be $\sim$50--100 km s$^{-1}$ by
comparing wavelengths of Ly$\alpha$ measured in spectra obtained with G160M and
G140L gratings. The parameters of Ly$\alpha$ emission line are shown in
Table\,\ref{tab4}. Equivalent widths
EW(Ly$\alpha$) of the Ly$\alpha$ emission line are high and peak separations
$V_{\rm sep}$ are low in seven out of eight galaxies indicating considerable
leakage of both the Ly$\alpha$ and LyC emission. An exception is J0238$+$0003.
Despite of the noisy spectrum (Fig.~\ref{fig5}c) two weak components of
the Ly$\alpha$ emission line with low EWs are definitely detected.
Characteristics of both components were derived by Gaussian fitting with
taking into account the continuum level and the noise. The high peak separation
$V_{\rm sep}$ implies that LyC leakage from J0238$+$0003 is unlikely
\citep[e.g. ][]{I18b}.

The Ly$\alpha$ escape fraction $f_{\rm esc}$(Ly$\alpha$) is derived using
equation
\begin{equation}
f_{\rm esc}({\rm Ly}\alpha)=\frac{1}{A}\left(\frac{I({\rm Ly}\alpha)}{I({\rm H}\beta)}\right),  \label{eq:fescla}
\end{equation}
where
\begin{equation}
A= \left(\frac{I({\rm Ly}\alpha)}{I({\rm H}\beta)}\right)_{\rm rec}.  \label{eq:rec}
\end{equation}
In Eq.~\ref{eq:fescla}, $I$(Ly$\alpha$) is the Ly$\alpha$ flux corrected for
Milky Way extinction, $I$(H$\beta$) is the H$\beta$ flux corrected for
extinction derived from the hydrogen Balmer decrement, and $A$ is the case B
recombination Ly$\alpha$/H$\beta$ flux ratio (Eq.~\ref{eq:rec}), which
depends on the electron temperature and electron number density in the
H~{\sc ii} region. According to \citet{SH95}, $A$ = 23.3 for physical conditions
in H~{\sc ii} regions of studied galaxies (Table~\ref{taba3}).

Characteristics of Ly$\alpha$ emission (Table~\ref{tab4}) closely follow
the relations considered by \citet{I24a} for low-redshift LyC and Ly$\alpha$
leaking galaxies. We also note that Ly$\alpha$ luminosities and equivalent
widths are very similar to that in luminous Ly$\alpha$ galaxies at the end of
reionisation \citep[e.g. ][]{Ni26}.

The observed G140L spectra (grey lines) and extrapolations to the UV range of
predicted intrinsic SEDs in the optical range (blue dash-dotted
lines) are shown in Fig.~\ref{fig6}.
Additionally, we include the attenuated
extrapolations of the intrinsic SEDs (black solid lines),
the same as those with $R(V)$ = 2.7 that are shown in Fig.\,\ref{fig4}
but with different flux and wavelength scales.

The photon count budget in the Lyman continuum is shown in
Table~\ref{taba4}. From the gross counts in the given rest frame wavelength
range we computed the source and the background counts with Poisson statistical
1$\sigma$ uncertainties \citep{FC98}. We also give
separate components of the background counts: dark current, scattered
light \citep{W16} and the diffuse FUV background \citep{Mu2014}.
The 1$\sigma$ uncertainties on the background components are computed
separately for the dark current \citep{Ma21} and the modelled
scattered light \citep{W16}. For the diffuse FUV background we
assume an uncertainty of 10\%. The background uncertainties are propagated
into the total background. The last column in the table is the
false-positive probability P($>$N|B) that the source counts are actually a
background fluctuation \citep{W16,I18a}.
The source counts (and the respective fluxes) also have an estimate of
the systematic error due to the background uncertainty (2nd pair of
uncertainties in Table~\ref{tab5}). The systematic error was estimated from
Monte Carlo simulations of the background adopting a Gaussian distribution
with a mean of the total background and a standard deviation of the
respective background uncertainty \citep{W16}.

The level of the observed LyC continuum is indicated by horizontal
red lines. The vertical dotted lines show the Lyman limit.
The LyC emission is securely detected in spectra of three galaxies,
J0131+2242, J1340+5448, J1638+3348,
and tentatively detected in spectra of four galaxies, J0021+3117,
J1247+3154, J1607+3217, J1658+3647 (solid red lines in Fig.~\ref{fig6} and
insets in the Figure). The galaxy J0238+0003 has an unreliable LyC
measurement because of possible contamination.
The measurements are summarised in Table\,\ref{tab5}.

\citet{I16a,I16b,I18a,I18b} used the ratio of the escaping
fluxes $I_{\rm esc}$ to the intrinsic fluxes $I_{\rm mod}$
of the Lyman continuum to derive $f_{\rm esc}$(LyC):
\begin{equation}
f_{\rm esc}({\rm LyC}) =\frac{I_{\rm esc}(\lambda)}{I_{\rm mod}(\lambda)}, 
\label{eq:fesc}
\end{equation}
where $\lambda$ is the mean wavelength of the range near 900\,\AA\ used for
averaging the LyC flux density (see Table~\ref{tab5}).
The wavelength windows were defined before inspection of the data. As
in previous work we chose 20\AA\ rest frame intervals. We measured the LyC flux
in the closest to 900\AA\ uncontaminated 20\AA\ window whenever possible. For
two targets (J0021+3117, J0131+2242; Fig.~\ref{fig6}) the Lyman limit is
adjacent to geocoronal Lya emission and measurements are potentially affected by
systematic errors of the scattered light model. Therefore, we used the
window 880-900\AA. For consistency, we adopted this window as well
for J1607+3217, J1638+3348 and J1658+3647. For the higher-redshift
targets of our sample, geocoronal Lya clobbers a large part of the LyC,
leaving the first uncontaminated wavelength windows at 850-870\AA\ 
(J1247+3154) and 840-860\AA\ (J1340+5448). Lastly, for J0238+0003 we chose
a very blue wavelength window (840-860\AA) to minimize potential scattered
light residuals, as the observations suffered from very strong geocoronal
contamination also during orbital shadow (see geocoronal residuals of
O~{\sc i} 1304\AA\ and O~{\sc i}] 1356\AA\ in Fig.~\ref{fig6}).

We iteratively derive the intrinsic fluxes 
$I_{\rm mod}$ and, correspondingly, the LyC escape fractions $f_{\rm esc}$(LyC) 
from simultaneous fitting of the SED in the optical range together with
observed equivalent widths of the H$\beta$ and H$\alpha$ emission lines.

Results of the $f_{\rm esc}$(LyC) measurements are shown in Table\,\ref{tab5}.
We obtain the escape fractions
$f_{\rm esc}$(LyC) in the range between $\sim$~4 and $\sim$~40 per cent in seven
out of the eight galaxies. However, $I_{\rm obs}$ and $f_{\rm esc}$(LyC)
are derived above 3$\sigma$ only in three galaxies, J0131+2242, J1340+5448, and
J1638+3348. The detection of LyC in other galaxies should be considered as
tentative.

The most surprising result is that the LyC flux in the eighth galaxy,
J0238$+$0003, is high, contrary to that expected
from the properties of the Ly$\alpha$ emission line. The Lyman
continuum in this galaxy is likely contaminated by emission of probable
faint diffuse foreground galaxy near the brightest knot of J0238$+$0003
(Fig.~\ref{fig2}). Surface brightness profile for J0238$+$0003 in the NUV range
(Fig.~\ref{fig3}) reveals that surface brightness from the diffuse galaxy is by
$\sim$\,1 mag weaker than in the brightest knot. On the other hand, the area 
with this emission is approximately two times larger than that for the
emission from the brightest knot, resulting in comparable total NUV fluxes from
the brightest knot and diffuse component. Assuming the same emission proportion
for the FUV range and that this diffuse emission emerges in the foreground
galaxy we may expect comparable fluxes of emission from foreground galaxy
in the LyC range and of non-ionising emission from J0238$+$0003
at $\lambda$ $>$ 912\AA\ (Fig.~\ref{fig6}c).
However, existing observational data are not sufficient to
confirm the presence of a foreground galaxy.
We conclude that $f_{\rm esc}$(LyC) for this galaxy is unreliable.

\section{Indirect determination of the LyC escape fraction} \label{sec:ind}

LyC emission in most low-$z$ galaxies is difficult directly to
detect because of its weakness. Furthermore, only {\sl HST} can be used for the
observation  of galaxies with $z$~$\sim$~0.3~--~0.4 in the LyC wavelength range.
As for high-$z$ galaxies during the epoch of reionisation, incomplete
ionisation of the intergalactic medium prevents direct detection of the
Lyman continuum. Therefore, indirect methods of the LyC leakage determination
for large samples of galaxies at low and high redshift are needed.
Until now only slightly above one hundred direct LyC
measurements of LyC emission in low-$z$ galaxies with various masses,
metallicities and ionisation conditions in their H~{\sc ii} regions
have been done. In the present paper we discuss the properties of Ly$\alpha$
and ionising emission in galaxies with the lowest possible oxygen abundances
found in the SDSS at redshifts of $\sim$~0.3~--~0.4.

Several indirect indicators have been proposed, which are based mainly
on observations of non-ionising continuum and strong emission lines in the UV
and optical ranges, as well as absorption lines of low-ionisation species.
These observations are easier to obtain compared to those of
LyC emission. They include such indicators as the Ly$\alpha$ profile, the slope
of the continuum in the UV range at wavelengths longer than that of the
Lyman limit, strong emission lines in the rest-frame optical range, their
intensities and equivalent widths, as well as their intensity ratios.
Some of these observations can be done with ground-based telescopes.

One of the most promising indicators is the Ly$\alpha$ emission line.
The dependence of $f_{\rm esc}$(Ly$\alpha$)
and $f_{\rm esc}$(LyC) on the oxygen abundance is shown in Figs.~\ref{fig7}a and
\ref{fig7}b. It is seen that our new objects considerably extend the range
of 12+log(O/H) to lower values filling the range with $\sim$ 7.5 -- 7.7.
We find that there is only weak dependence of both the
escape fractions $f_{\rm esc}$(Ly$\alpha$) and $f_{\rm esc}$(LyC) on oxygen
abundance (Figs.~\ref{fig7}a and \ref{fig7}b), although $f_{\rm esc}$(LyC)
in our lowest-metallicity galaxies (red symbols) are systematically higher
than that in galaxies e.g. from the LzLCS sample with
higher metallicities (green symbols). Thus, in general, galaxy metallicity is
not a reliable indicator of escaping Ly$\alpha$ and LyC emission.
We can only conclude that most metal-deficient galaxies are among the highest
LyC and Ly$\alpha$ leakers.

The Ly$\alpha$ escape fraction $f_{\rm esc}$(Ly$\alpha$), which is derived from
the Ly$\alpha$/H$\beta$ emission line ratio, can potentially be linked
with the LyC escape fraction despite the differences between mechanisms
controlling the escape of LyC and Ly$\alpha$ emission. The LyC photons
can efficiently be absorbed by neutral hydrogen
and/or dust. On the other hand, Ly$\alpha$ photons can be ceased only via
absorption by dust and via inefficient two-photon transitions. Thus,
the fraction of escaping Ly$\alpha$ photons is expected to be higher than
that of escaping LyC photons, in agreement with theoretical predictions
\citep{D15,JO13,NO14,N22,AMo26}. This is seen in Fig.\,\ref{fig7}c, where almost
all LyC leaking galaxies are located below the line of equal
escape fractions (black solid line). Our galaxies with the lowest metallicities
are expected to have lower dust content, and therefore higher $f_{\rm esc}$(LyC).
This is indeed the case, all our galaxies (red symbols) have systematically
higher $f_{\rm esc}$(LyC) compared to that for LyC leaking
galaxies with higher metallicities. There is a tendency for $f_{\rm esc}$(LyC)
to increase with increasing $f_{\rm esc}$(Ly$\alpha$), but with a large spread
\citep[see also e.g. ][]{I18b,I21a,F22b}. New data do not contradict this
conclusion. Somewhat apart is the galaxy J0238$+$0003
(crossed red symbol) with very high LyC flux and very low Ly$\alpha$ flux.
We suggest that the LyC wavelength range in the spectrum of this galaxy is
contaminated by emission from a foreground galaxy.

The shape of the Ly$\alpha$ profile provides one of the best indirect
indicators of the LyC leakage due to the fact that it depends on the column
density of the neutral hydrogen along the line of sight, which determines
the optical depth in both the Ly$\alpha$ emission line and the LyC continuum.
The separation between blue and red components of the Ly$\alpha$ line in
medium-resolution COS spectra is the measure of the neutral hydrogen optical
depth. Previously, \cite{V17} and \citet{I18b} found
a correlation of $f_{\rm esc}$(LyC) on the separation $V_{\rm sep}$ between
the peaks of the Ly$\alpha$ emission line in LyC leakers. This dependence
has been updated in the later paper by \citet{I21a}.
The new data also follow the relation discussed by
\citet{I18b} (see the solid line in Fig.\,\ref{fig7}d). However, we note
that all our data are slightly above the analytical approximation by
\citet{I18b}, perhaps, due to the low metallicity and likely low dust content.

There is only one
galaxy, J0238$+$0003 (crossed red symbol) that strongly deviates  from
the relation established for other galaxies and thus shows that escaping
LyC emission is totally unrelated to escaping Ly$\alpha$ emission.
This deviation can be understood adopting, for example, the presence of
foreground galaxy in the direction on J0238$+$0003.

There is a slight trend of decreasing $f_{\rm esc}$(LyC) with increasing
stellar mass in galaxies with both the detected LyC continuum and its upper
limit (Fig.\,\ref{fig7}e), in agreement with earlier studies
\citep[e.g. ][]{I21a}. New data in the
present paper are in agreement with the conclusion of only a weak
correlation between $f_{\rm esc}$(LyC) and $M_\star$. In fact, the average value
of $f_{\rm esc}$(LyC) in galaxies with detected LyC emission is above 10\%
in galaxies with $M_\star$ $<$ 10$^8$M$_\odot$, and drops down to $\sim$~5\% in
galaxies with $M_\star$~$\sim$~10$^{10}$M$_\odot$ (Fig.~\ref{fig7}e).

Another promising indicator is the O$_{32}$ =
[O~{\sc iii}]$\lambda$5007/[O~{\sc ii}]$\lambda$3727 ratio \citep{JO13,NO14}.
In general, the increase of this ratio is caused not only by decreasing
optical depth of the neutral hydrogen around the H~{\sc ii} region, but
also by increasing ionisation parameter and/or decreasing metallicity.
These effects are difficult to separate, but they all act in the same
direction.
The relation between $f_{\rm esc}$(LyC) and O$_{32}$ has been discussed
by \citet{F16}, \citet{I18b,I21a} and \citet{F22b}. Its updated version
from \citet{I25} is presented in
Fig.\,\ref{fig7}f, which shows a trend of increasing $f_{\rm esc}$(LyC)
with increasing of O$_{32}$, but with a substantial scatter. This scatter,
in part, can be caused by a variety of scenarios with leakage through channels
with low optical depth and their orientation relative to the
observer. A similar conclusion can be drawn from the \citet{F22b} data.
A high O$_{32}$ is the second best indicator among those considered
in this paper and it can be used for selection of the LyC
leaking candidates. However, it is less certain indicator of
high $f_{\rm esc}$(LyC) $\ga$ 10\% compared to the small separation
$V_{\rm sep}$ of the Ly$\alpha$ emission line \citep{I18b,Na20}. We note that
our extremely metal-poor LyC leakers are not the galaxies with extreme O$_{32}$
values.

\section{Conclusions}\label{sec:summary}

We present new {\sl HST} COS low- and medium-resolution spectra of
eight compact SFG in the redshift range
$z$ = 0.3012 -- 0.3780, with oxygen abundances of 7.49 -- 7.74, considerably
lower than the oxygen abundances of most LyC leaking galaxies studied so far.
We aim to obtain properties of leaking LyC and resolved Ly$\alpha$
emission and to study the dependence
of leaking Ly$\alpha$ and LyC emission on the oxygen abundances
along with other indirect indicators of escaping ionising radiation.
This study is an extension of the work
reported earlier in \citet{I16a,I16b,I18a,I18b,I21a,I22,I25}. Our main
results are summarised as follows:
\begin{itemize}
\item Metallicities of our new galaxies, ranging from 12 + logO/H
= 7.49 to 7.74, are considerably lower than those in the LyC leakers
studied by \citet{I16b,I18a,I18b,I21a,I22,I25}.
\item Emission of Lyman continuum is measured in seven out of the
eight galaxies with the escape fraction $f_{\rm esc}$(LyC) in the range
between 4.6\% and 41.7\%. These low metallicity, low dust content galaxies
appear to commonly emit ionising photons. This strongly links the escape of
ionising photons to the observations of faint galaxies in the early Universe.
However, LyC emission is securely detected at the $>$~3$\sigma$ level
only in three galaxies, J0131+2242, J1340+5448, and J1638+3348. The detection
of LyC emission in the other four galaxies, J0021+3117, J1247+3154, J1607+3217,
and J1658+3647, should be considered only as tentative.
The eighth galaxy, J0238+0003, is excluded from the reliable LyC results
because of possible contamination.
\item A Ly$\alpha$ emission line with two peaks is observed in the
spectra of all galaxies. The Ly$\alpha$ emission line in one galaxy,
J0238$+$0003, is very weak. Excluding this galaxy, our new observations support
an anti-correlation between $f_{\rm esc}$(LyC) and the peak velocity
separation $V_{\rm sep}$ of the Ly$\alpha$ profile, consistent with the finding
of \citet{I18b,I21a} and making $V_{\rm sep}$ the most robust indirect
indicator of Lyman continuum radiation leakage.
\item Other characteristics such as oxygen abundance 12+log(O/H), O$_{32}$ ratio,
escape fraction of the Ly$\alpha$ emission line $f_{\rm esc}$(Ly$\alpha$) and the
stellar mass $M_\star$ show weaker correlations with $f_{\rm esc}$(LyC),
with a high spread of values, in agreement with earlier studies by e.g.
\citet{I16b,I18a,I21a}, \citet{F22b}.
\item We study the relations between $f_{\rm esc}$(Ly$\alpha$) and 12+log(O/H),
and $f_{\rm esc}$(LyC) and 12+log(O/H). No correlation is found for both
relations.
\item A bright compact star-forming region superimposed on a
low-surface-brightness emission are seen in the COS near ultraviolet
(NUV) acquisition images of most of the galaxies (one image is
missing due to technical problems). The surface brightness at the
outskirts of our galaxies can be approximated by an exponential disc, with a
scale length of $\sim$~0.24~--~0.73 kpc. This is $\sim$ 2 -- 4 times lower than
the scale lengths of the LyC leakers observed by \citet{I16b,I18a,I18b},
but is similar to that in low-mass galaxies with $M_\star$~$\la$~10$^8$~M$_\odot$
by \citet{I21a}.
Part of this difference may be explained by smaller acquisition
exposures that are $\sim$ 2 shorter compared to those used by
\citet{I16b,I18a,I18b}, resulting in less deep images.
\end{itemize}

\begin{acknowledgements}
Based on observations made with the NASA/ESA {\sl Hubble Space Telescope}, 
obtained from the data archive at the Space Telescope Science Institute. 
Support for TXT and JC was provided by NASA through grant number HST-GO-17109
from the Space Telescope Science Institute. STScI is operated by the
Association of Universities for Research in Astronomy,
Inc. under NASA contract NAS 5-26555. 
YI, DS, and NG acknowledge support from the joint Ukrainian-Swiss project
No. 224866.
Funding for SDSS-III has been provided by the Alfred P. Sloan Foundation, 
the Participating Institutions, the National Science Foundation, and the U.S. 
Department of Energy Office of Science. The SDSS-III website is 
http://www.sdss3.org/. SDSS-III is managed by the Astrophysical Research 
Consortium for the Participating Institutions of the SDSS-III Collaboration. 
{\sl GALEX} is a NASA mission  managed  by  the  Jet  Propulsion  Laboratory.
This research has made use of the NASA/IPAC Extragalactic Database (NED) which 
is operated by the Jet  Propulsion  Laboratory,  California  Institute  of  
Technology,  under  contract with the National Aeronautics and Space 
Administration.
\end{acknowledgements}

\begin{appendix}
\onecolumn

\section{Tables}

  \begin{table*}[ht!]
  \caption{Apparent AB magnitudes with errors in parentheses compiled
from the SDSS and {\sl GALEX} databases and apparent Vega magnitudes from
the {\sl WISE} database
\label{taba1}}
\begin{tabular}{lccccccccccccc} \hline
Name&\multicolumn{5}{c}{SDSS}
&&\multicolumn{2}{c}{\sl GALEX}&&\multicolumn{4}{c}{{\sl WISE}} \\ 
    &\multicolumn{1}{c}{$u$}&\multicolumn{1}{c}{$g$}&\multicolumn{1}{c}{$r$}&\multicolumn{1}{c}{$i$}&\multicolumn{1}{c}{$z$}&&FUV&NUV&&\multicolumn{1}{c}{$W1$}&\multicolumn{1}{c}{$W2$}&\multicolumn{1}{c}{$W3$}&\multicolumn{1}{c}{$W4$}
\\
    &\multicolumn{1}{c}{(err)}&\multicolumn{1}{c}{(err)}&\multicolumn{1}{c}{(err)}&\multicolumn{1}{c}{(err)}&\multicolumn{1}{c}{(err)}&&(err)&(err)&&\multicolumn{1}{c}{(err)}&\multicolumn{1}{c}{(err)}&\multicolumn{1}{c}{(err)}&\multicolumn{1}{c}{(err)} \\
\hline
J0021$+$3117& 22.06& 21.93& 20.81& 21.98& 21.19&&  ... & 21.63&&\multicolumn{1}{c}{ ... }&\multicolumn{1}{c}{ ... }&\multicolumn{1}{c}{ ... }&\multicolumn{1}{c}{ ... } \\
            &(0.23)&(0.07)&(0.05)&(0.19)&(0.37)&& (...)&(0.43)&&\multicolumn{1}{c}{(...)}&\multicolumn{1}{c}{(...)}&\multicolumn{1}{c}{(...)}&\multicolumn{1}{c}{(...)} \\
J0131$+$2242& 22.96& 22.38& 21.65& 22.74& 21.88&& 21.64&  ... &&\multicolumn{1}{c}{ ... }&\multicolumn{1}{c}{ ... }&\multicolumn{1}{c}{ ... }&\multicolumn{1}{c}{ ... } \\
            &(0.34)&(0.10)&(0.07)&(0.25)&(0.48)&&(0.35)& (...)&&\multicolumn{1}{c}{(...)}&\multicolumn{1}{c}{(...)}&\multicolumn{1}{c}{(...)}&\multicolumn{1}{c}{(...)} \\
J0238$+$0003& 22.29& 21.89& 21.63& 21.53& 21.79&&  ... & 21.71&&\multicolumn{1}{c}{17.05}&\multicolumn{1}{c}{ ... }&\multicolumn{1}{c}{ ... }&\multicolumn{1}{c}{ ... } \\
            &(0.18)&(0.06)&(0.06)&(0.08)&(0.41)&& (...)&(0.37)&&\multicolumn{1}{c}{(0.10)}&\multicolumn{1}{c}{(...)}&\multicolumn{1}{c}{(...)}&\multicolumn{1}{c}{(...)} \\
J1247$+$3154& 22.21& 21.66& 21.23& 22.12& 20.77&& 22.03& 21.87&&\multicolumn{1}{c}{ ... }&\multicolumn{1}{c}{ ... }&\multicolumn{1}{c}{ ... }&\multicolumn{1}{c}{ ... } \\
            &(0.24)&(0.06)&(0.06)&(0.21)&(0.24)&&(0.34)&(0.28)&&\multicolumn{1}{c}{(...)}&\multicolumn{1}{c}{(...)}&\multicolumn{1}{c}{(...)}&\multicolumn{1}{c}{(...)} \\
J1340$+$5448& 22.35& 22.29& 22.17& 21.92& 21.30&&  ... &  ... &&\multicolumn{1}{c}{ ... }&\multicolumn{1}{c}{ ... }&\multicolumn{1}{c}{ ... }&\multicolumn{1}{c}{ ... } \\
            &(0.21)&(0.10)&(0.13)&(0.14)&(0.29)&& (...)& (...)&&\multicolumn{1}{c}{(...)}&\multicolumn{1}{c}{(...)}&\multicolumn{1}{c}{(...)}&\multicolumn{1}{c}{(...)} \\
J1607$+$3217& 20.96& 20.55& 19.51& 20.92& 19.81&& 20.88& 20.69&&\multicolumn{1}{c}{17.09}&\multicolumn{1}{c}{15.50}&\multicolumn{1}{c}{11.41}&\multicolumn{1}{c}{ 8.57} \\
            &(0.08)&(0.02)&(0.01)&(0.06)&(0.08)&&(0.28)&(0.15)&&\multicolumn{1}{c}{(0.10)}&\multicolumn{1}{c}{(0.09)}&\multicolumn{1}{c}{(0.17)}&\multicolumn{1}{c}{(0.30)} \\
J1638$+$3348& 22.23& 22.16& 21.38& 22.45& 21.39&& 22.30&  ... &&\multicolumn{1}{c}{ ... }&\multicolumn{1}{c}{ ... }&\multicolumn{1}{c}{ ... }&\multicolumn{1}{c}{ ... } \\
            &(0.18)&(0.07)&(0.06)&(0.20)&(0.32)&&(0.19)& (...)&&\multicolumn{1}{c}{(...)}&\multicolumn{1}{c}{(...)}&\multicolumn{1}{c}{(...)}&\multicolumn{1}{c}{(...)} \\
J1658$+$3647& 22.05& 21.65& 21.09& 22.16& 21.12&& 21.14& 21.36&&\multicolumn{1}{c}{ ... }&\multicolumn{1}{c}{ ... }&\multicolumn{1}{c}{ ... }&\multicolumn{1}{c}{ ... } \\
            &(0.19)&(0.06)&(0.05)&(0.22)&(0.36)&&(0.30)&(0.29)&&\multicolumn{1}{c}{(...)}&\multicolumn{1}{c}{(...)}&\multicolumn{1}{c}{(...)}&\multicolumn{1}{c}{(...)} \\
\hline
\end{tabular}

  \end{table*}

  \begin{table*}[ht!]
  \caption{Extinction-corrected fluxes and rest-frame equivalent widths of
the emission lines in SDSS spectra
\label{taba2}}
\begin{tabular}{lcrrrrrrrr} \hline
 & &\multicolumn{8}{c}{Galaxy}\\ \cline{3-10}
Line &\multicolumn{1}{c}{$\lambda$}&\multicolumn{2}{c}{J0021$+$3117}& \multicolumn{2}{c}{J0131$+$2242}& \multicolumn{2}{c}{J0238$+$0003}& \multicolumn{2}{c}{J1247$+$3154}\\ \cline{3-10}
     &&\multicolumn{1}{c}{$I$$^{\rm a}$}&\multicolumn{1}{c}{EW$^{\rm b}$}&\multicolumn{1}{c}{$I$$^{\rm a}$}&\multicolumn{1}{c}{EW$^{\rm b}$}&\multicolumn{1}{c}{$I$$^{\rm a}$}&\multicolumn{1}{c}{EW$^{\rm b}$}&\multicolumn{1}{c}{$I$$^{\rm a}$}&\multicolumn{1}{c}{EW$^{\rm b}$}\\
\hline
Mg~{\sc ii}          &2796&  21.6$\pm$3.0&  18$\pm$2 &      ...    & ...        & 19.0$\pm$2.1&  15$\pm$2 &      ...    & ...       \\
Mg~{\sc ii}          &2803&   8.6$\pm$1.9&   6$\pm$1 &      ...    & ...        &  9.0$\pm$1.5&   7$\pm$1 &      ...    & ...       \\
$[$O~{\sc ii}$]$     &3727&  55.5$\pm$5.2&  97$\pm$10 & 55.5$\pm$4.3&  75$\pm$5 & 90.8$\pm$4.8& 159$\pm$7 &103.1$\pm$6.6& 103$\pm$10\\
H12                  &3750&       ...    & ...        &      ...    & ...       &      ...    & ...       &  6.3$\pm$1.8&   9$\pm$1 \\
H11                  &3771&       ...    & ...        &      ...    & ...       &      ...    & ...       &  6.5$\pm$1.9&   9$\pm$1 \\
H10                  &3798&   6.2$\pm$2.1&  17$\pm$2  &      ...    & ...       &      ...    & ...       &  5.7$\pm$1.8&   7$\pm$1 \\
H9                   &3836&  11.8$\pm$2.6&  33$\pm$3  &  6.5$\pm$1.8&   9$\pm$3 &  8.0$\pm$1.8&  16$\pm$3 &  8.0$\pm$2.0&  12$\pm$1 \\
$[$Ne~{\sc iii}$]$   &3869&  47.1$\pm$4.8& 108$\pm$11 & 38.8$\pm$3.5&  75$\pm$7 & 40.1$\pm$3.1&  63$\pm$4 & 46.7$\pm$4.4&  43$\pm$4 \\
H8+He~{\sc i}        &3889&  20.6$\pm$3.3&  48$\pm$5  & 25.5$\pm$3.0&  40$\pm$5 & 20.8$\pm$2.4&  48$\pm$5 & 24.5$\pm$3.2&  33$\pm$3 \\
H7+$[$Ne~{\sc iii}$]$&3969&  30.9$\pm$4.1&  60$\pm$6  & 29.0$\pm$3.2&  54$\pm$6 & 28.8$\pm$2.8&  65$\pm$4 & 29.4$\pm$3.5&  38$\pm$4 \\
H$\delta$            &4101&  24.6$\pm$4.0&  27$\pm$3  & 25.6$\pm$3.3&  33$\pm$4 & 26.0$\pm$2.6&  68$\pm$6 & 30.9$\pm$3.6&  39$\pm$4 \\
H$\gamma$            &4340&  51.2$\pm$5.2& 116$\pm$12 & 45.9$\pm$3.9& 106$\pm$9 & 49.5$\pm$3.5&  98$\pm$7 & 55.5$\pm$4.7& 105$\pm$11\\
$[$O~{\sc iii}$]$    &4363&  19.7$\pm$3.2&  53$\pm$5  & 18.9$\pm$2.8&  37$\pm$5 & 17.3$\pm$2.2&  33$\pm$4 & 15.8$\pm$2.7&  30$\pm$3 \\
He~{\sc i}           &4471&       ...    & ...        &  7.2$\pm$1.8&  11$\pm$3 &  5.2$\pm$1.6&   8$\pm$2 &      ...    & ...       \\
H$\beta$             &4861& 100.0$\pm$7.5& 181$\pm$18 &100.0$\pm$6.2& 135$\pm$8 &100.0$\pm$5.2& 228$\pm$10&100.0$\pm$6.5& 161$\pm$16\\
$[$O~{\sc iii}$]$    &4959& 204.4$\pm$11.& 327$\pm$33 &179.0$\pm$8.6& 396$\pm$16&185.7$\pm$7.4& 351$\pm$12&181.8$\pm$9.0& 329$\pm$33\\
$[$O~{\sc iii}$]$    &5007& 588.1$\pm$21.&1030$\pm$99 &517.5$\pm$17.& 782$\pm$29&581.7$\pm$17.&1089$\pm$20&563.3$\pm$19.&1090$\pm$99\\
He~{\sc i}           &5876&  10.9$\pm$2.7&  48$\pm$5  & 12.0$\pm$2.6&  31$\pm$7 &  8.5$\pm$1.9&  20$\pm$4 &  8.7$\pm$2.3&  30$\pm$3 \\
H$\alpha$            &6563& 307.7$\pm$16.&2605$\pm$261&304.5$\pm$14.&1273$\pm$45&265.7$\pm$11.& 619$\pm$19&270.1$\pm$13.& 588$\pm$59\\
$[$N~{\sc ii}$]$     &6583&   5.4$\pm$1.9&  71$\pm$7  &      ...    & ...       &  9.5$\pm$1.9&  22$\pm$4 &      ...    & ...       \\
$[$S~{\sc ii}$]$     &6717&       ...    & ...        &      ...    & ...       &  6.1$\pm$1.6&  16$\pm$4 &  8.5$\pm$2.2&  76$\pm$8 \\
$[$S~{\sc ii}$]$     &6731&       ...    & ...        &      ...    & ...       &  4.4$\pm$1.6&  11$\pm$4 &  6.9$\pm$2.1&  51$\pm$5 \\
$C$(H$\beta$)$^{\rm c}$  &&\multicolumn{2}{c}{0.100$\pm$0.063}&\multicolumn{2}{c}{0.080$\pm$0.055}&\multicolumn{2}{c}{0.110$\pm$0.048}&\multicolumn{2}{c}{0.100$\pm$0.060}\\
$F$(H$\beta$)$^{\rm d}$     &&\multicolumn{2}{c}{8.1$\pm$0.6}&\multicolumn{2}{c}{5.5$\pm$0.3}&\multicolumn{2}{c}{7.5$\pm$0.3}&\multicolumn{2}{c}{10.7$\pm$0.7}\\
\hline
 & &\multicolumn{8}{c}{Galaxy}\\ \cline{3-10}
Line &\multicolumn{1}{c}{$\lambda$}&\multicolumn{2}{c}{J1340$+$5448}& \multicolumn{2}{c}{J1607$+$3217}& \multicolumn{2}{c}{J1638$+$3348}& \multicolumn{2}{c}{J1658$+$3647} \\ \cline{3-10}
     &&\multicolumn{1}{c}{$I$$^{\rm a}$}&\multicolumn{1}{c}{EW$^{\rm b}$}&\multicolumn{1}{c}{$I$$^{\rm a}$}&\multicolumn{1}{c}{EW$^{\rm b}$}&\multicolumn{1}{c}{$I$$^{\rm a}$}&\multicolumn{1}{c}{EW$^{\rm b}$}&\multicolumn{1}{c}{$I$$^{\rm a}$}&\multicolumn{1}{c}{EW$^{\rm b}$}\\
\hline
Mg~{\sc ii}          &2796&  26.4$\pm$3.9&  29$\pm$3  & 12.7$\pm$1.4&   9$\pm$1 &      ...    & ...        & 36.2$\pm$4.3&  18$\pm$2 \\
Mg~{\sc ii}          &2803&  12.9$\pm$2.7&  15$\pm$2  & 12.2$\pm$1.4&   8$\pm$1 &      ...    & ...        &  7.8$\pm$1.8&   8$\pm$1 \\
$[$O~{\sc ii}$]$     &3727&  74.8$\pm$6.9& 147$\pm$15 & 77.5$\pm$3.6&  67$\pm$7 & 49.9$\pm$5.4&  54$\pm$5  & 90.2$\pm$7.3&  76$\pm$8 \\
H12                  &3750&       ...    & ...        &  4.9$\pm$1.5&   5$\pm$1 &      ...    & ...        &      ...    & ...       \\
H11                  &3771&       ...    & ...        &  5.9$\pm$1.5&   6$\pm$1 &      ...    & ...        &      ...    & ...       \\
H10                  &3798&       ...    & ...        &  7.0$\pm$1.4&   8$\pm$1 &      ...    & ...        & 10.8$\pm$3.3&   9$\pm$1 \\
H9                   &3836&   8.9$\pm$2.8&  15$\pm$1  &  9.5$\pm$1.5&  12$\pm$1 &  6.5$\pm$2.6&   5$\pm$1  & 10.3$\pm$3.4&  10$\pm$1 \\
$[$Ne~{\sc iii}$]$   &3869&  49.6$\pm$5.5& 116$\pm$12 & 40.0$\pm$2.4&  56$\pm$6 & 39.3$\pm$5.0&  37$\pm$4  & 42.4$\pm$5.1&  32$\pm$3 \\
H8+He~{\sc i}        &3889&  19.8$\pm$3.6&  40$\pm$4  & 18.7$\pm$1.8&  25$\pm$3 & 21.5$\pm$3.5&  25$\pm$3  & 21.1$\pm$4.4&  14$\pm$1 \\
H7+$[$Ne~{\sc iii}$]$&3969&  31.1$\pm$4.7&  53$\pm$5  & 30.5$\pm$2.3&  42$\pm$4 & 24.2$\pm$4.0&  37$\pm$4  & 29.7$\pm$5.0&  21$\pm$2 \\
H$\delta$            &4101&  24.2$\pm$4.1&  57$\pm$6  & 26.4$\pm$2.1&  39$\pm$4 & 24.5$\pm$4.0&  32$\pm$3  & 26.3$\pm$4.7&  19$\pm$2 \\
H$\gamma$            &4340&  46.2$\pm$5.6& 165$\pm$16 & 47.9$\pm$2.8&  87$\pm$9 & 44.7$\pm$5.4&  49$\pm$5  & 48.5$\pm$5.8&  45$\pm$5 \\
$[$O~{\sc iii}$]$    &4363&  22.7$\pm$3.9&  98$\pm$10 & 12.2$\pm$1.4&  23$\pm$2 & 18.8$\pm$3.5&  26$\pm$3  & 15.4$\pm$3.5&  17$\pm$2 \\
He~{\sc i}           &4471&   4.8$\pm$2.1&  15$\pm$1  &  3.7$\pm$1.0&   8$\pm$1 &      ...    & ...        &      ...    & ...       \\
H$\beta$             &4861& 100.0$\pm$8.6& 329$\pm$32 &100.0$\pm$4.2& 215$\pm$21&100.0$\pm$8.1& 263$\pm$26 &100.0$\pm$12.& 134$\pm$13\\
$[$O~{\sc iii}$]$    &4959& 199.8$\pm$13.& 813$\pm$71 &143.3$\pm$3.7& 312$\pm$31&162.0$\pm$11.& 370$\pm$36 &199.7$\pm$12.& 266$\pm$27\\
$[$O~{\sc iii}$]$    &5007& 595.1$\pm$25.&1997$\pm$130&429.0$\pm$9.0& 937$\pm$94&531.4$\pm$22.&1206$\pm$121&590.6$\pm$23.& 789$\pm$79\\
He~{\sc i}           &5876&   8.8$\pm$2.9&  50$\pm$5  & 11.6$\pm$1.5&  35$\pm$4 & 10.2$\pm$3.2&  16$\pm$2  & 12.6$\pm$3.5&  19$\pm$2 \\
H$\alpha$            &6563& 252.1$\pm$17.& 831$\pm$83 &255.4$\pm$8.8& 754$\pm$75&334.1$\pm$19.& 802$\pm$80 &351.7$\pm$20.& 346$\pm$35\\
$[$N~{\sc ii}$]$     &6583&       ...    & ...        &  8.1$\pm$1.6&  14$\pm$1 &      ...    & ...        &  4.7$\pm$2.2&   9$\pm$1 \\
$[$S~{\sc ii}$]$     &6717&       ...    & ...        &  7.1$\pm$1.4&  26$\pm$3 &      ...    & ...        &  9.8$\pm$3.2&  16$\pm$2 \\
$[$S~{\sc ii}$]$     &6731&       ...    & ...        &  5.1$\pm$1.2&  18$\pm$2 &      ...    & ...        &  9.0$\pm$3.2&  15$\pm$2 \\
$C$(H$\beta$)$^{\rm c}$  &&\multicolumn{2}{c}{0.020$\pm$0.082}&\multicolumn{2}{c}{0.080$\pm$0.0.041}&\multicolumn{2}{c}{0.100$\pm$0.069}&\multicolumn{2}{c}{0.050$\pm$0.066}\\
$F$(H$\beta$)$^{\rm d}$     &&\multicolumn{2}{c}{6.1$\pm$0.5}&\multicolumn{2}{c}{30.3$\pm$1.1}&\multicolumn{2}{c}{6.0$\pm$0.5}&\multicolumn{2}{c}{6.8$\pm$0.5}\\
\hline
  \end{tabular}
\tablefoot{$^{\rm a}$$I$=100$\times$$I$($\lambda$)/$I$(H$\beta$), where $I$($\lambda$) and $I$(H$\beta$) are fluxes of emission lines, corrected for both the 
Milky Way and internal extinction. $^{\rm b}$Rest frame equivalent width in \AA. $^{\rm c}$Extinction coefficient. $^{\rm d}$Observed flux in 10$^{-16}$ erg s$^{-1}$ cm$^{-2}$.}
\end{table*}

  \begin{table*}[ht!]
  \caption{Electron temperatures, electron number densities and 
element abundances in H~{\sc ii} regions \label{taba3}}
  \begin{tabular}{lcccc} \hline
Galaxy &J0021$+$3117 &J0131$+$2242 &J0238$+$0003 &J1247$+$3154 \\ \hline
$T_{\rm e}$ ($[$O {\sc iii}$]$), K      & 19810$\pm$1990       & 20880$\pm$2010       & 18750$\pm$1410     & 18120$\pm$1720\\
$T_{\rm e}$ ($[$O {\sc ii}$]$), K       & 15600$\pm$1450       & 15640$\pm$1380       & 15440$\pm$1080     & 15300$\pm$1350 \\
$N_{\rm e}$ ($[$S {\sc ii}$]$), cm$^{-3}$& 10$\pm$10$^{\rm a}$       &  10$\pm$10$^{\rm a}$  &       34$^{+373}_{-13}$   &   212$^{+417}_{-153}$    \\ \\
O$^+$/H$^+$$\times$10$^{5}$             &0.45$\pm$0.05        &0.44$\pm$0.04        &0.75$\pm$0.06       &0.90$\pm$0.09 \\
O$^{2+}$/H$^+$$\times$10$^{5}$          &3.46$\pm$0.12        &2.74$\pm$0.10        &3.76$\pm$0.09      &3.94$\pm$0.11  \\
O/H$\times$10$^{5}$                    &3.91$\pm$0.13        &3.18$\pm$0.11        &4.52$\pm$0.11      &4.84$\pm$0.14  \\
12+log O/H                             &7.59$\pm$0.01        &7.50$\pm$0.01        &7.65$\pm$0.01      &7.68$\pm$0.01  \\ \\
N$^+$/H$^+$$\times$10$^{6}$             &0.37$\pm$0.14       &       ...            &0.66$\pm$0.16      &     ...       \\
ICF(N)$^{\rm b}$                        &    8.08            &        ...           &5.68               &     ...       \\
N/H$\times$10$^{6}$                     &2.98$\pm$1.22       &       ...           &3.73$\pm$0.86      &     ...        \\
log N/O                                &~$-$1.12$\pm$0.18~~\, &      ...          &~$-$1.08$\pm$0.10~~\,&    ...       \\ \\
Ne$^{2+}$/H$^+$$\times$10$^{5}$          &0.61$\pm$0.07        &0.45$\pm$0.04        &0.59$\pm$0.05       &0.75$\pm$0.08 \\
ICF(Ne)$^{\rm b}$                       &1.05                 &1.06                 &1.07                &1.08           \\
Ne/H$\times$10$^{5}$                    &0.64$\pm$0.07        &0.47$\pm$0.05        &0.63$\pm$0.05       &0.81$\pm$0.09 \\
log Ne/O                               &~$-$0.79$\pm$0.05~~\, &~$-$0.83$\pm$0.04~~\, &~$-$0.85$\pm$0.04~~\,&~$-$0.78$\pm$0.05~~\, \\ \\
Mg$^{+}$/H$^+$$\times$10$^{6}$          &0.17$\pm$0.03        &        ...          & 0.16$\pm$0.02      &      ...     \\
ICF(Mg)$^{\rm b}$                        &15.56                &       ...          &      11.43       &        ...       \\
Mg/H$\times$10$^{6}$                    &2.64$\pm$0.50       &        ...          & 1.88$\pm$0.27     &        ...      \\
log Mg/O                               &~$-$1.17$\pm$0.08~~\, &      ...           &~$-$1.38$\pm$0.06~~\, &     ...      \\
\hline
Galaxy &J1340$+$5448 &J1607$+$3217 &J1638$+$3348&J1658$+$3647 \\ \cline{1-5}
$T_{\rm e}$ ($[$O {\sc iii}$]$), K      & 21600$\pm$2510       & 18190$\pm$1200       & 20910$\pm$2560 & 17300$\pm$2100 \\
$T_{\rm e}$ ($[$O {\sc ii}$]$), K       & 15600$\pm$1650       & 15310$\pm$940         & 15640$\pm$1750 & 15050$\pm$1700 \\
$N_{\rm e}$ ($[$S {\sc ii}$]$), cm$^{-3}$&  10$\pm$10$^{\rm a}$       &   28$^{+229}_{-7}$  &        10$\pm$10$^{\rm a}$    &   450$^{+835}_{-304}$ \\ \\
O$^+$/H$^+$$\times$10$^{5}$             &0.60$\pm$0.07        &0.66$\pm$0.05        &0.40$\pm$0.05  &0.85$\pm$0.11   \\
O$^{2+}$/H$^+$$\times$10$^{5}$          &2.92$\pm$0.14        &3.00$\pm$0.05        &2.72$\pm$0.12   &4.63$\pm$0.16   \\
O/H$\times$10$^{5}$                    &3.52$\pm$0.16        &3.71$\pm$0.07        &3.12$\pm$0.13   &5.48$\pm$0.19   \\
12+log O/H                             &7.55$\pm$0.02        &7.57$\pm$0.01        &7.49$\pm$0.02   &7.74$\pm$0.02   \\ \\
N$^+$/H$^+$$\times$10$^{6}$             &        ...        &0.57$\pm$0.12        &       ...         &0.34$\pm$0.17     \\
ICF(N)$^{\rm b}$                        &        ...         &5.35                   &     ...         &6.07           \\
N/H$\times$10$^{6}$                     &       ...         &3.08$\pm$0.69        &       ...         &2.08$\pm$1.12      \\
log N/O                                &        ...         &~$-$1.08$\pm$0.10~~\, &     ...        &~$-$1.42$\pm$0.23~~\,\\ \\
Ne$^{2+}$/H$^+$$\times$10$^{5}$          &0.53$\pm$0.06        &0.64$\pm$0.05        &0.45$\pm$0.06   &0.77$\pm$0.11   \\
ICF(Ne)$^{\rm b}$                       &1.07                 &1.08                 &1.05             &1.07             \\
Ne/H$\times$10$^{5}$                    &0.57$\pm$0.07        &0.69$\pm$0.05        &0.47$\pm$0.06   &0.82$\pm$0.12    \\
log Ne/O                               &~$-$0.79$\pm$0.05~~\, &~$-$0.73$\pm$0.03~~\, &~$-$0.82$\pm$0.06~~\, &~$-$0.83$\pm$0.07~~\, \\ \\
Mg$^{+}$/H$^+$$\times$10$^{6}$          &0.22$\pm$0.05         &  0.15$\pm$0.02     &         ...        &  0.29$\pm$0.07     \\
ICF(Mg)$^{\rm b}$                        &     11.21           &     10.12          &         ...        &     12.620         \\
Mg/H$\times$10$^{6}$                    &2.48$\pm$0.52        & 1.54$\pm$0.19      &         ...         & 3.55$\pm$0.78      \\
log Mg/O                               &~$-$1.15$\pm$0.09~~\, &~$-$1.38$\pm$0.06~~\, &      ...         &~$-$1.19$\pm$0.10~~\,\\
\hline
\end{tabular}
\tablefoot{$^{\rm a}$Adopted value of $N_{\rm e}$ in galaxy with non-detected [S~{\sc ii}] 6717, 6731\AA\ emission lines. $^{\rm b}$Ionisation correction factor.}
  \end{table*}

  \begin{table*}[ht!]
  \caption{Photon counts in spectra of selected galaxies obtained with the G140L grating \label{taba4}}
\centering
\begin{tabular}{lccrcrccc} \hline
&&\multicolumn{6}{c}{Counts}& \\ \cline{3-8}
Name&		$\lambda$$^{\rm a}$, \AA&Gross$^{\rm b}$&	Background$^{\rm c}$&      Source$^{\rm d}$&\multicolumn{1}{c}{Dark$^{\rm e}$}&Scatter$^{\rm f}$&FUV backgr$^{\rm g}$&	P($>$N|B)$^{\rm h}$ \\ \hline
\vspace{0.1cm}
J0021+3117&	880-900&	120&	95.28$\pm$3.75&	24.72$^{+11.32}_{-10.67}$&	62.60$\pm$2.79&	28.10$\pm$2.47&	4.59$\pm$0.46&	6.268e-03 \\
\vspace{0.1cm}
J0131+2242&	880-900&	167&	90.65$\pm$4.21&	76.35$^{+13.32}_{-12.67}$&	43.22$\pm$1.83&	39.93$\pm$3.76&	7.51$\pm$0.75&	2.502e-13 \\
\vspace{0.1cm}
J0238+0003&	840-860&	207&	116.41$\pm$4.57&90.59$^{+14.83}_{-14.16}$&	76.47$\pm$2.95&	35.83$\pm$3.51&	4.12$\pm$0.41&	1.371e-14 \\
\vspace{0.1cm}
J1247+3154&	850-870&	107&	81.39$\pm$3.67&	25.61$^{+10.82}_{-10.17}$&	48.08$\pm$2.01&	30.70$\pm$3.11&	2.60$\pm$0.26&	2.762e-03 \\
\vspace{0.1cm}
J1340+5448&	840-860&	201&	122.68$\pm$6.95&78.32$^{+14.82}_{-14.17}$&	44.62$\pm$1.63&	72.35$\pm$6.82&	5.70$\pm$0.57&	3.507e-11 \\
\vspace{0.1cm}
J1607+3217&	880-900&	124&	102.31$\pm$4.58&21.69$^{+11.82}_{-11.17}$&	59.93$\pm$2.07&	37.17$\pm$4.15&	5.21$\pm$0.52&	1.632e-02 \\
\vspace{0.1cm}
J1638+3348&	880-900&	228&	170.42$\pm$6.52&57.58$^{+15.82}_{-15.81}$&	110.32$\pm$3.82&52.32$\pm$5.32&	7.77$\pm$0.78&	1.123e-05 \\
\vspace{0.1cm}
J1658+3647&	880-900&	117&	95.62$\pm$3.44&	21.38$^{+11.32}_{-10.67}$&	60.84$\pm$1.60&	29.51$\pm$3.06&	5.26$\pm$0.53&	1.478e-02 \\
\hline
\end{tabular}
\tablefoot{$^{\rm a}$Rest frame wavelength range for computing the source and the background counts. $^{\rm b}$Gross = Background + Source. $^{\rm c}$ and $^{\rm d}$ The source and background counts with Poisson statistical 1$\sigma$ uncertainties. Background = Dark + Scatter + FUV backgr. $^{\rm e}$Dark current. $^{\rm f}$Modelled scattered light. $^{\rm g}$Diffuse FUV background. $^{\rm h}$False-positive probability P($>$N|B) that the source counts are actually a background fluctuation.}

  \end{table*}

\end{appendix}


\begin{thebibliography}{}

\bibitem[Ade et al.(2014)]{P14} Ade, P. A. R., Aghanim, N., Armitage-Caplan, C.,
et al. 2014, \aap, 571, A16

\bibitem[Aller(1984)]{A84} Aller, L. H. 1984, Physics of Thermal
Gaseous Nebulae. (Dordrecht: Reidel)

\bibitem[Almada Monter et al.(2026)]{AMo26} Almada Monter, S., Gronke, M.,
\& Chang, S.-J. 2026, \mnras, 547, 1

\bibitem[Arellano-C\'ordova et al.(2022)]{AC22} Arellano-C\'ordova, K. Z.,
Berg, D. A., Chisholm, J., et al. 2022, \apj, 940, L23

\bibitem[Baldwin et al.(1981)]{BPT81} 
Baldwin, J. A., Phillips, M. M., \& Terlevich, R. 1981, \pasp, 93, 5

\bibitem[Bian et al.(2017)]{B17} Bian, F., Fan, X., McGreer, I., Cai, Z., \& Jiang L. 2017, \apj, 837, 12

\bibitem[Borthakur et al.(2014)]{B14} Borthakur, S., Heckman, T. M., 
Leitherer, C., \& Overzier, R. A. 2014, \sci, 346, 216

\bibitem[Bouwens et al.(2015)]{B15a} Bouwens, R. J., Illingworth, G. D., 
Oesch, P. A., et al. 2015, \apj, 811, 140

\bibitem[Boyett et al.(2024)]{Bo24} Boyett, K., Bunker, A. J., Curtis-Lake, E.,
et al. 2024, \mnras, 535, 1796

\bibitem[Cameron et al.(2023)]{Ca23} Cameron, A. J., Saxena, A.,
Bunker, A. J., et al. 2023, \aap, 677, A115

\bibitem[Cardelli et al.(1989)]{C89} 
Cardelli, J. A., Clayton, G. C., \& Mathis, J. S. 1989, \apj, 345, 245

\bibitem[Chisholm et al.(2017)]{C17} Chisholm, J., Orlitov\'a, I., 
Schaerer, D., et al. 2017, \aap, 605, A67

\bibitem[Chisholm et al.(2018)]{Ch18} Chisholm, J., Gazagnes, S., Schaerer, D.,
et al. 2018, \aap, 616, 30

\bibitem[Chisholm et al.(2022)]{Ch22} Chisholm, J., Saldana-Lopez, A.,
Flury, S., et al. 2022, \mnras, 517, 5104

\bibitem[Curti et al.(2024)]{Cu24} Curti, M., Maiolino, R., Curtis-Lake, E.,
et al. 2024, \aap, 684, A75

\bibitem[de Barros et al.(2016)]{B16} de Barros, S., Vanzella, E., Amorín, R.,
et al. 2016, \aap, 585, A51

\bibitem[Dressler et al.(2015)]{D15} Dressler, A., Henry, A., Martin, C. L., 
et al. 2015, \apj, 806, 19

\bibitem[Endsley et al.(2023)]{E23} Endsley, R., Stark, D. P., Whitler, L.,
et al. 2023, \mnras, 524, 2312

\bibitem[Faisst(2016)]{F16} Faisst, A. L. 2016, \apj, 829, 99

\bibitem[Feldman \& Cousins(1998)]{FC98} Feldman, G. J., \& Cousins, R. D.
1998, \prd, 57, 3873

\bibitem[Finkelstein et al.(2019)]{Fi19} Finkelstein, S. L., D'Aloisio, A.,
Paardekooper, J.-P., et al. 2019, \apj, 879, 36

\bibitem[Flury et al.(2022a)]{F22a} Flury, S. R., Jaskot, A. E.,
Ferguson, H. C., et al. 2022a, \apjs, 260, 1

\bibitem[Flury et al.(2022b)]{F22b} Flury, S. R., Jaskot, A. E.,
Ferguson, H. C., et al. 2022b, \apj, 930, 126

\bibitem[Gazagnes et al.(2018)]{Ga18} Gazagnes, S., Chisholm, J., Schaerer, D., 
et al. 2018, \aap, 616, 29

\bibitem[Gazagnes et al.(2020)]{Ga20} Gazagnes, S., Chisholm, J., Schaerer, D., 
Verhamme, A., \& Izotov, Y. 2020, \aap, 639, 85 

\bibitem[Girardi et al.(2000)]{G00} Girardi, L., Bressan, A., Bertelli, G. 
\& Chiosi, C. 2000, \aaps, 141, 371

\bibitem[Grazian et al.(2016)]{Gr16} Grazian, A. Giallongo, E., Gerbasi, R.,
et al. 2016, \aap, 585, A48 

\bibitem[Guseva et al.(2024)]{G24} Guseva, N. G., Thuan, T. X., \&
Izotov, Y. I. 2024, \mnras, 527, 3932

\bibitem[Heintz et al.(2023)]{He23} Heintz, K. E., Brammer, G. B.,
Gim\'enez-Arteaga, C., et al. 2023, \nata, 7, 1517

\bibitem[Henry et al.(2015)]{H15} Henry, A., Scarlata, C., Martin, C. S., \&
Erb, D. 2015, \apj, 809, 19

\bibitem[Hsiao et al.(2026)]{H26} Hsiao, T. Y.-Y., Chisholm, J., Berg, D. A.,
et al. 2026, \apj, submitted; arXiv:2605.06770

\bibitem[Inoue et al.(2014)]{Inoue14} Inoue, A.~K., Shimizu, I., Iwata, I.
\& Tanaka, M. 2014, \mnras, 442, 1805 

\bibitem[Izotov et al.(2006)]{I06} Izotov, Y. I., Stasi\'nska, G., Meynet, G.,
Guseva, N. G., \& Thuan, T. X. 2006, \aap, 448, 955

\bibitem[Izotov et al.(2016a)]{I16a} Izotov, Y. I., Orlitov\'a, I.,
Schaerer, D., et al. 2016a, \nat, 529, 178

\bibitem[Izotov et al.(2016b)]{I16b} Izotov, Y. I., Schaerer, D., Thuan, T. X., 
et al. 2016b, \mnras, 461, 3683

\bibitem[Izotov et al.(2017)]{I17} Izotov, Y. I., 
Thuan, T. X., \& Guseva, N. G. 2017, \mnras, 471, 548

\bibitem[Izotov et al.(2018a)]{I18a} Izotov, Y. I., Schaerer, D.,
Worseck, G., et al. 2018a, \mnras, 474, 4514

\bibitem[Izotov et al.(2018b)]{I18b} Izotov, Y. I., Worseck, G., Schaerer, D.,
et al. 2018b, \mnras, 478, 4851

\bibitem[Izotov et al.(2018c)]{I18c} Izotov, Y. I., Thuan, T. X., Guseva, N. G.
\& Liss, S. E. 2018c, \mnras, 473, 1956

\bibitem[Izotov et al.(2020)]{I20} Izotov, Y. I., Schaerer, D., Worseck, G., 
et al. 2020, \mnras, 491, 468

\bibitem[Izotov et al.(2021)]{I21a} Izotov, Y. I., Worseck, G., Schaerer, D., 
et al. 2021, \mnras, 503, 1734

\bibitem[Izotov et al.(2022)]{I22} Izotov, Y. I., Chisholm, J., Worseck, G.,
et al. 2022, \mnras, 515, 2864

\bibitem[Izotov et al.(2024)]{I24a} Izotov, Y. I., Thuan, T. X.,
Guseva, N. G., et al. 2024a, \mnras, 527, 281

\bibitem[Izotov et al.(2024)]{I24b} Izotov, Y. I., Thuan, T. X., \&
Guseva, N. G. 2024b, \mnras, 527, 3486

\bibitem[Izotov et al.(2025)]{I25} Izotov, Y. I., Schaerer, D., Worseck, G.,
et al. 2025, \aap, 704, A19

\bibitem[Jaskot \& Oey(2013)]{JO13} Jaskot, A. E., \& Oey, M. S.
2013, \apj, 766, 91

\bibitem[Jaskot \& Oey(2014)]{JO14} Jaskot, A. E., \& Oey, M. S.
2014, \apj, 791, L19

\bibitem[Jecmen et al.(2026)]{Je26} Jecmen, M. C., Chisholm, J., Atek, H.,
  et al. 2026, \apj, 1001, 234

\bibitem[Kauffmann et al.(2003)]{K03} Kauffmann, G., Heckman, T. M.,
Tremonti, C., et al. 2003, \mnras, 341, 33

\bibitem[Kunth \& \"Ostlin(2000)]{KO00} Kunth, D., \& \"Ostlin, G. 2000,
\aapr, 10, 1

\bibitem[Lejeune et al.(1997)]{L97} Lejeune ,T., 
Buser, R., \& Cuisinier, F. 1997, \aaps, 125, 229

\bibitem[Lin et al.(2024)]{L24} Yu-Heng Lin, Y.-H., Scarlata, C., Williams, H.,
et al. 2024, \mnras, 527, 4173

\bibitem[Makan et al.(2021)]{Ma21} Makan, K., Worseck, G., Davies, F. B.,
et al. 2021, \apj, 912, 38

\bibitem[Mathis(1990)]{Ma90} Mathis, J. S. 1990, \araa, 28, 10
  
\bibitem[Me\v{s}tric et al.(2025)]{Mes25} Me\v{s}tric, U., Vanzella, E.,
Beckett, A., et al. 2025, \aap, 698, A203 

\bibitem[Murthy(2014)]{Mu2014} Murthy, J. 2014, \apjs, 213, 32

\bibitem[Naidu et al.(2020)]{N20} Naidu, R. P., Tacchella, S., Mason, C. A., 
et al. 2020, \apj, 892, 109

\bibitem[Naidu et al.(2022)]{N22} Naidu, R. P., Matthee, J., Oesch, P. A.,
et al. 2022, \mnras, 510, 4582

\bibitem[Nakajima \& Ouchi(2014)]{NO14} Nakajima, K., \& Ouchi, M.
2014, \mnras, 442, 900

\bibitem[Nakajima et al.(2020)]{Na20} Nakajima, K., Ellis, R. S.,
Robertson, B. E., Tang, M., \& Stark, D. P. 2020, \apj, 889, 161 

\bibitem[Nakajima et al.(2023)]{Na23} Nakajima, K., Ouchi, M., Isobe, Y., et al.
2014, \apjs, 269, 33

\bibitem[Ning et al.(2026)]{Ni26} Ning, Y., Cai, Z., Jiang, L., et al.
2026, \apj, 1009, L4

\bibitem[Ouchi et al.(2009)]{O09} Ouchi, M., Mobasher, B., Shimasaku, K., et al.
2009, \apj, 706, 1136

\bibitem[Rivera-Thorsen et al.(2019)]{RT19} Rivera-Thorsen, T. E., Dahle, H.,
Chisholm, J., et al. 2019, \sci, 366, 738

\bibitem[Roberts-Borsani et al.(2024)]{RB24} Roberts-Borsani, G.,
Treu, T., Shapley, A., et al. 2024, \apj, 976, 193

\bibitem[Robertson et al.(2013)]{R13} Robertson, B. E., Furlanetto, S. R.,
Schneider, E., et al. 2013, \apj, 768, 71

\bibitem[Saldana-Lopez et al.(2022)]{SL22} Saldana-Lopez, A., Schaerer, D.,
Chisholm, J., et al. 2022, \aap, 663, 59

\bibitem[Salpeter(1955)]{S55} Salpeter, E. E. 1955, ApJ, 121, 161

\bibitem[Sanders et al.(2023)]{San23} Sanders, R. L., Shapley, A. E.,
Topping, M. W., Reddy, N. A., \& Brammer, G. B. 2023, \apj, 955, 54

\bibitem[Saxena et al.(2024)]{Sa24} Saxena, A., Bunker, A. J., Jones, G. C.,
et al. 2024, \aap, 684, A84

\bibitem[Schaerer et al.(2022)]{S22} Schaerer, D., Marques-Chaves, R.,
Barrufet, L., et al. 2022, \aap, 665, L4

\bibitem[Schmutz et al.(1992)]{S92} 
Schmutz, W., Leitherer, C., \& Gruenwald, R. 1992, \pasp, 104, 1164

\bibitem[Steidel et al.(2018)]{St18} Steidel, C. C., Bogosavljevi\'c, M., 
Shapley, A. E., et al. 2018, \apj, 869, 123

\bibitem[Storey \& Hummer(1995)]{SH95} Storey, P. J., \& Hummer, D. G.
1995, \mnras, 272, 41

\bibitem[Vanzella et al.(2010)]{V10} Vanzella, E., Giavalisco, M.,
Inoue, A. K., et al. 2010, \apj, 725, 1011 

\bibitem[Vanzella et al.(2012)]{V12} Vanzella, E., Guo, Y. Giavalisco, M.,
et al. 2012, \apj, 751, 70 

\bibitem[Vanzella et al.(2018)]{Va18} Vanzella, E., Nonino, M., Cupani, G.,
et al. 2018, \mnras, 476, L15 

\bibitem[Verhamme et al.(2015)]{V15} Verhamme, A., Orlitov\'a, I., 
Schaerer, D., \& Hayes, M. 2015, \aap, 578, A7

\bibitem[Verhamme et al.(2017)]{V17} Verhamme, A., Orlitov\'a, I., Schaerer, D.,
et al. 2017, \aap, 597, A13

\bibitem[Wang et al.(2026)]{W26} Wang, S., Wang, X., Zhou, H.,
et al. 2026, \apj, 1006, L17

\bibitem[Wise et al.(2014)]{W14} Wise, J.~H., Demchenko, 
V.~G., Halicek, M.~T., et al. 2014, \mnras, 442, 2560 

\bibitem[Worseck et al.(2016)]{W16} Worseck, G., Prochaska, J. X.,
Hennawi, J. F., \& McQuinn, M. 2016, \apj, 825, 144

\bibitem[Wright(2006)]{W06} Wright, E. L. 2006, \pasp, 118, 1711

\bibitem[Xu et al.(2022)]{Xu22} Xu, X., Henry, A., Heckman, T., et al. 2022,
\apj, 933, 202

\bibitem[Yang et al.(2017)]{Y17} Yang, H., Malhotra, S., Gronke, M., et al.
2017, \apj, 844, 171

\bibitem[Zhu et al.(2026)]{Zhu26} Zhu, S., Zheng, Z.-Y., Bian, F., et al.
2026, \apj, 1005, L3


\end{thebibliography}
\end{document}